\documentclass[10pt,conference]{IEEEtran}

\usepackage{cite}
\usepackage[T1]{fontenc}
\usepackage{graphicx}
\usepackage{amssymb}
\usepackage{amsmath}
\usepackage{amsthm}
\usepackage{subfigure}
\usepackage{booktabs} %\toprule \midrule \bottomrule
\usepackage{multirow}
\usepackage{microtype}
\usepackage{balance}
\usepackage{xcolor}
\usepackage[hidelinks]{hyperref}
\usepackage{mathdots}
\usepackage{epstopdf}
\usepackage{algorithm}
\usepackage{algpseudocode}
\usepackage{setspace}
\usepackage{footmisc}
\usepackage{tikz}
\usepackage{circledsteps}
\usepackage{bbm}
\usepackage{xfrac}
\usepackage{dblfloatfix}

\usepackage{amssymb}
\usepackage{amsfonts}
\usepackage{mathrsfs}
\usepackage{xspace}
\usepackage{bm}
\usepackage{upgreek}

\newcommand{\safemath}[2]{\newcommand{#1}{\ensuremath{#2}\xspace}}

\safemath{\bma}{\mathbf{a}}
\safemath{\bmb}{\mathbf{b}}
\safemath{\bmc}{\mathbf{c}}
\safemath{\bmd}{\mathbf{d}}
\safemath{\bme}{\mathbf{e}}
\safemath{\bmf}{\mathbf{f}}
\safemath{\bmg}{\mathbf{g}}
\safemath{\bmh}{\mathbf{h}}
\safemath{\bmi}{\mathbf{i}}
\safemath{\bmj}{\mathbf{j}}
\safemath{\bmk}{\mathbf{k}}
\safemath{\bml}{\mathbf{l}}
\safemath{\bmm}{\mathbf{m}}
\safemath{\bmn}{\mathbf{n}}
\safemath{\bmo}{\mathbf{o}}
\safemath{\bmp}{\mathbf{p}}
\safemath{\bmq}{\mathbf{q}}
\safemath{\bmr}{\mathbf{r}}
\safemath{\bms}{\mathbf{s}}
\safemath{\bmt}{\mathbf{t}}
\safemath{\bmu}{\mathbf{u}}
\safemath{\bmv}{\mathbf{v}}
\safemath{\bmw}{\mathbf{w}}
\safemath{\bmx}{\mathbf{x}}
\safemath{\bmy}{\mathbf{y}}
\safemath{\bmz}{\mathbf{z}}
\safemath{\bmzero}{\mathbf{0}}
\safemath{\bmone}{\mathbf{1}}

\bmdefine{\biad}{a}
\bmdefine{\bibd}{b}
\bmdefine{\bicd}{c}
\bmdefine{\bidd}{d}
\bmdefine{\bied}{e}
\bmdefine{\bifd}{f}
\bmdefine{\bigd}{g}
\bmdefine{\bihd}{h}
\bmdefine{\biid}{i}
\bmdefine{\bijd}{j}
\bmdefine{\bikd}{k}
\bmdefine{\bild}{l}
\bmdefine{\bimd}{m}
\bmdefine{\bind}{n}
\bmdefine{\biod}{o}
\bmdefine{\bipd}{p}
\bmdefine{\biqd}{q}
\bmdefine{\bird}{r}
\bmdefine{\bisd}{s}
\bmdefine{\bitd}{t}
\bmdefine{\biud}{u}
\bmdefine{\bivd}{v}
\bmdefine{\biwd}{w}
\bmdefine{\bixd}{x}
\bmdefine{\biyd}{y}
\bmdefine{\bizd}{z}

\bmdefine{\bixid}{\xi}
\bmdefine{\bilambdad}{\lambda}
\bmdefine{\bimud}{\mu}
\bmdefine{\bithetad}{\theta}
\bmdefine{\biphid}{\phi}
\bmdefine{\bideltad}{\delta}

\safemath{\bmia}{\biad}
\safemath{\bmib}{\bibd}
\safemath{\bmic}{\bicd}
\safemath{\bmid}{\bidd}
\safemath{\bmie}{\bied}
\safemath{\bmif}{\bifd}
\safemath{\bmig}{\bigd}
\safemath{\bmih}{\bihd}
\safemath{\bmii}{\biid}
\safemath{\bmij}{\bijd}
\safemath{\bmik}{\bikd}
\safemath{\bmil}{\bild}
\safemath{\bmim}{\bimd}
\safemath{\bmin}{\bind}
\safemath{\bmio}{\biod}
\safemath{\bmip}{\bipd}
\safemath{\bmiq}{\biqd}
\safemath{\bmir}{\bird}
\safemath{\bmis}{\bisd}
\safemath{\bmit}{\bitd}
\safemath{\bmiu}{\biud}
\safemath{\bmiv}{\bivd}
\safemath{\bmiw}{\biwd}
\safemath{\bmix}{\bixd}
\safemath{\bmiy}{\biyd}
\safemath{\bmiz}{\bizd}

\safemath{\bmxi}{\bixid}
\safemath{\bmlambda}{\bilambdad}
\safemath{\bmmu}{\bimud}
\safemath{\bmtheta}{\bithetad}
\safemath{\bmphi}{\biphid}
\safemath{\bmdelta}{\bideltad}

\safemath{\bA}{\mathbf{A}}
\safemath{\bB}{\mathbf{B}}
\safemath{\bC}{\mathbf{C}}
\safemath{\bD}{\mathbf{D}}
\safemath{\bE}{\mathbf{E}}
\safemath{\bF}{\mathbf{F}}
\safemath{\bG}{\mathbf{G}}
\safemath{\bH}{\mathbf{H}}
\safemath{\bI}{\mathbf{I}}
\safemath{\bJ}{\mathbf{J}}
\safemath{\bK}{\mathbf{K}}
\safemath{\bL}{\mathbf{L}}
\safemath{\bM}{\mathbf{M}}
\safemath{\bN}{\mathbf{N}}
\safemath{\bO}{\mathbf{O}}
\safemath{\bP}{\mathbf{P}}
\safemath{\bQ}{\mathbf{Q}}
\safemath{\bR}{\mathbf{R}}
\safemath{\bS}{\mathbf{S}}
\safemath{\bT}{\mathbf{T}}
\safemath{\bU}{\mathbf{U}}
\safemath{\bV}{\mathbf{V}}
\safemath{\bW}{\mathbf{W}}
\safemath{\bX}{\mathbf{X}}
\safemath{\bY}{\mathbf{Y}}
\safemath{\bZ}{\mathbf{Z}}

\safemath{\bZero}{\mathbf{0}}
\safemath{\bOne}{\mathbf{1}}
\safemath{\bDelta}{\mathbf{\Delta}}
\safemath{\bLambda}{\mathbf{\UpLambda}}
\safemath{\bPhi}{\mathbf{\Upphi}}
\safemath{\bSigma}{\mathbf{\Upsigma}}
\safemath{\bOmega}{\mathbf{\Upomega}}
\safemath{\bTheta}{\mathbf{\Uptheta}}

\bmdefine{\biAd}{A}
\bmdefine{\biBd}{B}
\bmdefine{\biCd}{C}
\bmdefine{\biDd}{D}
\bmdefine{\biEd}{E}
\bmdefine{\biFd}{F}
\bmdefine{\biGd}{G}
\bmdefine{\biHd}{H}
\bmdefine{\biId}{I}
\bmdefine{\biJd}{J}
\bmdefine{\biKd}{K}
\bmdefine{\biLd}{L}
\bmdefine{\biMd}{M}
\bmdefine{\biOd}{N}
\bmdefine{\biPd}{O}
\bmdefine{\biQd}{P}
\bmdefine{\biRd}{R}
\bmdefine{\biSd}{S}
\bmdefine{\biTd}{T}
\bmdefine{\biUd}{U}
\bmdefine{\biVd}{V}
\bmdefine{\biWd}{W}
\bmdefine{\biXd}{X}
\bmdefine{\biYd}{Y}
\bmdefine{\biZd}{Z}

\bmdefine{\biDelta}{\Delta}
\bmdefine{\biLambda}{\Lambda}
\bmdefine{\biPhi}{\Phi}
\bmdefine{\biSigma}{\Sigma}
\bmdefine{\biOmega}{\Omega}
\bmdefine{\biTheta}{\Theta}

\safemath{\bimA}{\biAd}
\safemath{\bimB}{\biBd}
\safemath{\bimC}{\biCd}
\safemath{\bimD}{\biDd}
\safemath{\bimE}{\biEd}
\safemath{\bimF}{\biFd}
\safemath{\bimG}{\biGd}
\safemath{\bimH}{\biHd}
\safemath{\bimI}{\biId}
\safemath{\bimJ}{\biJd}
\safemath{\bimK}{\biKd}
\safemath{\bimL}{\biLd}
\safemath{\bimM}{\biMd}
\safemath{\bimN}{\biNd}
\safemath{\bimO}{\biOd}
\safemath{\bimP}{\biPd}
\safemath{\bimQ}{\biQd}
\safemath{\bimR}{\biRd}
\safemath{\bimS}{\biSd}
\safemath{\bimT}{\biTd}
\safemath{\bimU}{\biUd}
\safemath{\bimV}{\biVd}
\safemath{\bimW}{\biWd}
\safemath{\bimX}{\biXd}
\safemath{\bimY}{\biYd}
\safemath{\bimZ}{\biZd}

\safemath{\bimDelta}{\biDelta}
\safemath{\bimLambda}{\biLambda}
\safemath{\bimPhi}{\biPhi}
\safemath{\bimSigma}{\biSigma}
\safemath{\bimOmega}{\biOmega}
\safemath{\bimTheta}{\biTheta}

\safemath{\setA}{\mathcal{A}}
\safemath{\setB}{\mathcal{B}}
\safemath{\setC}{\mathcal{C}}
\safemath{\setD}{\mathcal{D}}
\safemath{\setE}{\mathcal{E}}
\safemath{\setF}{\mathcal{F}}
\safemath{\setG}{\mathcal{G}}
\safemath{\setH}{\mathcal{H}}
\safemath{\setI}{\mathcal{I}}
\safemath{\setJ}{\mathcal{J}}
\safemath{\setK}{\mathcal{K}}
\safemath{\setL}{\mathcal{L}}
\safemath{\setM}{\mathcal{M}}
\safemath{\setN}{\mathcal{N}}
\safemath{\setO}{\mathcal{O}}
\safemath{\setP}{\mathcal{P}}
\safemath{\setQ}{\mathcal{Q}}
\safemath{\setR}{\mathcal{R}}
\safemath{\setS}{\mathcal{S}}
\safemath{\setT}{\mathcal{T}}
\safemath{\setU}{\mathcal{U}}
\safemath{\setV}{\mathcal{V}}
\safemath{\setW}{\mathcal{W}}
\safemath{\setX}{\mathcal{X}}
\safemath{\setY}{\mathcal{Y}}
\safemath{\setZ}{\mathcal{Z}}
\safemath{\emptySet}{\varnothing}

\safemath{\colA}{\mathscr{A}}
\safemath{\colB}{\mathscr{B}}
\safemath{\colC}{\mathscr{C}}
\safemath{\colD}{\mathscr{D}}
\safemath{\colE}{\mathscr{E}}
\safemath{\colF}{\mathscr{F}}
\safemath{\colG}{\mathscr{G}}
\safemath{\colH}{\mathscr{H}}
\safemath{\colI}{\mathscr{I}}
\safemath{\colJ}{\mathscr{J}}
\safemath{\colK}{\mathscr{K}}
\safemath{\colL}{\mathscr{L}}
\safemath{\colM}{\mathscr{M}}
\safemath{\colN}{\mathscr{N}}
\safemath{\colO}{\mathscr{O}}
\safemath{\colP}{\mathscr{P}}
\safemath{\colQ}{\mathscr{Q}}
\safemath{\colR}{\mathscr{R}}
\safemath{\colS}{\mathscr{S}}
\safemath{\colT}{\mathscr{T}}
\safemath{\colU}{\mathscr{U}}
\safemath{\colV}{\mathscr{V}}
\safemath{\colW}{\mathscr{W}}
\safemath{\colX}{\mathscr{X}}
\safemath{\colY}{\mathscr{Y}}
\safemath{\colZ}{\mathscr{Z}}

\safemath{\opA}{\mathbb{A}}
\safemath{\opB}{\mathbb{B}}
\safemath{\opC}{\mathbb{C}}
\safemath{\opD}{\mathbb{D}}
\safemath{\opE}{\mathbb{E}}
\safemath{\opF}{\mathbb{F}}
\safemath{\opG}{\mathbb{G}}
\safemath{\opH}{\mathbb{H}}
\safemath{\opI}{\mathbb{I}}
\safemath{\opJ}{\mathbb{J}}
\safemath{\opK}{\mathbb{K}}
\safemath{\opL}{\mathbb{L}}
\safemath{\opM}{\mathbb{M}}
\safemath{\opN}{\mathbb{N}}
\safemath{\opO}{\mathbb{O}}
\safemath{\opP}{\mathbb{P}}
\safemath{\opQ}{\mathbb{Q}}
\safemath{\opR}{\mathbb{R}}
\safemath{\opS}{\mathbb{S}}
\safemath{\opT}{\mathbb{T}}
\safemath{\opU}{\mathbb{U}}
\safemath{\opV}{\mathbb{V}}
\safemath{\opW}{\mathbb{W}}
\safemath{\opX}{\mathbb{X}}
\safemath{\opY}{\mathbb{Y}}
\safemath{\opZ}{\mathbb{Z}}
\safemath{\opZero}{\mathbb{O}}
\safemath{\identityop}{\opI}

\safemath{\veca}{\bma}
\safemath{\vecb}{\bmb}
\safemath{\vecc}{\bmc}
\safemath{\vecd}{\bmd}
\safemath{\vece}{\bme}
\safemath{\vecf}{\bmf}
\safemath{\vecg}{\bmg}
\safemath{\vech}{\bmh}
\safemath{\veci}{\bmi}
\safemath{\vecj}{\bmj}
\safemath{\veck}{\bmk}
\safemath{\vecl}{\bml}
\safemath{\vecm}{\bmm}
\safemath{\vecn}{\bmn}
\safemath{\veco}{\bmo}
\safemath{\vecp}{\bmp}
\safemath{\vecq}{\bmq}
\safemath{\vecr}{\bmr}
\safemath{\vecs}{\bms}
\safemath{\vect}{\bmt}
\safemath{\vecu}{\bmu}
\safemath{\vecv}{\bmv}
\safemath{\vecw}{\bmw}
\safemath{\vecx}{\bmx}
\safemath{\vecy}{\bmy}
\safemath{\vecz}{\bmz}

\safemath{\veczero}{\bmzero}
\safemath{\vecone}{\bmone}
\safemath{\vecxi}{\bmxi}
\safemath{\veclambda}{\bmlambda}
\safemath{\vecmu}{\bmmu}
\safemath{\vectheta}{\bmtheta}
\safemath{\vecphi}{\bmphi}
\safemath{\vecdelta}{\bmdelta}

\safemath{\matA}{\bA}
\safemath{\matB}{\bB}
\safemath{\matC}{\bC}
\safemath{\matD}{\bD}
\safemath{\matE}{\bE}
\safemath{\matF}{\bF}
\safemath{\matG}{\bG}
\safemath{\matH}{\bH}
\safemath{\matI}{\bI}
\safemath{\matJ}{\bJ}
\safemath{\matK}{\bK}
\safemath{\matL}{\bL}
\safemath{\matM}{\bM}
\safemath{\matN}{\bN}
\safemath{\matO}{\bO}
\safemath{\matP}{\bP}
\safemath{\matQ}{\bQ}
\safemath{\matR}{\bR}
\safemath{\matS}{\bS}
\safemath{\matT}{\bT}
\safemath{\matU}{\bU}
\safemath{\matV}{\bV}
\safemath{\matW}{\bW}
\safemath{\matX}{\bX}
\safemath{\matY}{\bY}
\safemath{\matZ}{\bZ}
\safemath{\matzero}{\bmzero}

\safemath{\matDelta}{\bDelta}
\safemath{\matLambda}{\bLambda}
\safemath{\matPhi}{\bPhi}
\safemath{\matSigma}{\bSigma}
\safemath{\matOmega}{\bOmega}
\safemath{\matTheta}{\bTheta}

\safemath{\matidentity}{\matI}
\safemath{\matone}{\matO}

\safemath{\rnda}{A}
\safemath{\rndb}{B}
\safemath{\rndc}{C}
\safemath{\rndd}{D}
\safemath{\rnde}{E}
\safemath{\rndf}{F}
\safemath{\rndg}{G}
\safemath{\rndh}{H}
\safemath{\rndi}{I}
\safemath{\rndj}{J}
\safemath{\rndk}{K}
\safemath{\rndl}{L}
\safemath{\rndm}{M}
\safemath{\rndn}{N}
\safemath{\rndo}{O}
\safemath{\rndp}{P}
\safemath{\rndq}{Q}
\safemath{\rndr}{R}
\safemath{\rnds}{S}
\safemath{\rndt}{T}
\safemath{\rndu}{U}
\safemath{\rndv}{V}
\safemath{\rndw}{W}
\safemath{\rndx}{X}
\safemath{\rndy}{Y}
\safemath{\rndz}{Z}

\safemath{\rveca}{\bimA}
\safemath{\rvecb}{\bimB}
\safemath{\rvecc}{\bimC}
\safemath{\rvecd}{\bimD}
\safemath{\rvece}{\bimE}
\safemath{\rvecf}{\bimF}
\safemath{\rvecg}{\bimG}
\safemath{\rvech}{\bimH}
\safemath{\rveci}{\bimI}
\safemath{\rvecj}{\bimJ}
\safemath{\rveck}{\bimK}
\safemath{\rvecl}{\bimL}
\safemath{\rvecm}{\bimM}
\safemath{\rvecn}{\bimN}
\safemath{\rveco}{\bomO}
\safemath{\rvecp}{\bimP}
\safemath{\rvecq}{\bimQ}
\safemath{\rvecr}{\bimR}
\safemath{\rvecs}{\bimS}
\safemath{\rvect}{\bimT}
\safemath{\rvecu}{\bimU}
\safemath{\rvecv}{\bimV}
\safemath{\rvecw}{\bimW}
\safemath{\rvecx}{\bimX}
\safemath{\rvecy}{\bimY}
\safemath{\rvecz}{\bimZ}

\safemath{\rvecxi}{\bmxi}
\safemath{\rveclambda}{\bmlambda}
\safemath{\rvecmu}{\bmmu}
\safemath{\rvectheta}{\bmtheta}
\safemath{\rvecphi}{\bmphi}

\safemath{\rmatA}{\bimA}
\safemath{\rmatB}{\bimB}
\safemath{\rmatC}{\bimC}
\safemath{\rmatD}{\bimD}
\safemath{\rmatE}{\bimE}
\safemath{\rmatF}{\bimF}
\safemath{\rmatG}{\bimG}
\safemath{\rmatH}{\bimH}
\safemath{\rmatI}{\bimI}
\safemath{\rmatJ}{\bimJ}
\safemath{\rmatK}{\bimK}
\safemath{\rmatL}{\bimL}
\safemath{\rmatM}{\bimM}
\safemath{\rmatN}{\bimN}
\safemath{\rmatO}{\bimO}
\safemath{\rmatP}{\bimP}
\safemath{\rmatQ}{\bimQ}
\safemath{\rmatR}{\bimR}
\safemath{\rmatS}{\bimS}
\safemath{\rmatT}{\bimT}
\safemath{\rmatU}{\bimU}
\safemath{\rmatV}{\bimV}
\safemath{\rmatW}{\bimW}
\safemath{\rmatX}{\bimX}
\safemath{\rmatY}{\bimY}
\safemath{\rmatZ}{\bimZ}

\safemath{\rmatDelta}{\bimDelta}
\safemath{\rmatLambda}{\bimLambda}
\safemath{\rmatPhi}{\bimPhi}
\safemath{\rmatSigma}{\bimSigma}
\safemath{\rmatOmega}{\bimOmega}
\safemath{\rmatTheta}{\bimTheta}

\usepackage{amssymb}
\usepackage{amsfonts}
\usepackage{mathrsfs}
\usepackage{xspace}
\usepackage{bm}
\usepackage{fancyref}
\usepackage{textcomp}

\usepackage{multirow}
\usepackage{stmaryrd}

\newenvironment{textbmatrix}{	\setlength{\arraycolsep}{2.5pt}%
								\big[\begin{matrix}}{\end{matrix}\big]%
								\raisebox{0.08ex}{\vphantom{M}}}

\def\be{\begin{equation}}
\def\ee{\end{equation}}
\def\een{\nonumber \end{equation}}
\def\mat{\begin{bmatrix}}
\def\emat{\end{bmatrix}}
\def\btm{\begin{textbmatrix}}
\def\etm{\end{textbmatrix}}

\def\ba#1\ea{\begin{align}#1\end{align}}
\def\bas#1\eas{\begin{align*}#1\end{align*}}
\def\bs#1\es{\begin{split}#1\end{split}}
\def\bg#1\eg{\begin{gather}#1\end{gather}}
\def\bml#1\eml{\begin{multline}#1\end{multline}}
\def\bi#1\ei{\begin{itemize}#1\end{itemize}}

\newcommand{\lefto}{\mathopen{}\left}

\DeclareMathOperator{\Prob}{\opP}			% probability of an event
\DeclareMathOperator{\Exop}{\opE}			% expectation operator
\DeclareMathOperator{\erf}{erf}				% error function
\DeclareMathOperator{\erfc}{erfc}			% complementary error function
\newcommand{\orth}{\perp}					% orthogonal
\newcommand{\Ex}[2]{\ensuremath{\Exop_{#1}\lefto[#2\right]}} 	% expectation
\newcommand{\tp}[1]{\ensuremath{#1^{T}}} 		% transpose
\newcommand{\herm}[1]{\ensuremath{#1^{H}}} 	% hermitian transpose
\newcommand{\inv}[1]{\ensuremath{#1^{-1}}} 	% inverse
\safemath{\dirac}{\delta}					% Dirac delta
\safemath{\krond}{\dirac}					% Kronecker delta
\safemath{\upto}{\uparrow}
\safemath{\downto}{\downarrow}
\safemath{\iu}{j}							% imaginary unit
\safemath{\ev}{\lambda}						% eigenvalue
\safemath{\hilseqspace}{l^{2}}				% Hilbert sequence space
\newcommand{\banachfunspace}[1]{\setL^{#1}}	% Banach function space
\safemath{\hilfunspace}{\banachfunspace{2}}	% Hilbert function space
\safemath{\SNR}{\textit{SNR}} 				% signal to noise ratio
\safemath{\PAR}{\textit{PAR}} 				% signal to noise ratio
\safemath{\No}{N_0}							% noise spectral density
\safemath{\Es}{E_s}							% energy per symbol
\safemath{\Eb}{E_b}							% energy per bit
\safemath{\EbNo}{\frac{\Eb}{\No}}
\safemath{\EsNo}{\frac{\Es}{\No}}

\DeclareMathOperator{\CHop}{\ensuremath{\opH}} % channel operator
\safemath{\tvir}{\rndh_{\CHop}}				% time-varying impulse response
\safemath{\tvtf}{\rndl_{\CHop}}				% 	-''- transfer function
\safemath{\spf}{\rnds_{\CHop}}				% spreading function
\safemath{\bff}{H_{\CHop}}					% bi-freuqency function

\safemath{\ircf}{r_{h}}						% impulse response correlation fn.
\safemath{\tftvcf}{r_{s}}					% scattering function
\safemath{\tfcf}{r_{l}}						% time-frequency correlation fn.
\safemath{\bfcf}{r_{H}}						% bi-frequency correlation fn.

\safemath{\tcorr}{c_h}						% time-correlation function
\safemath{\scf}{c_{s}}						% spreading function
\safemath{\tfcorr}{c_{l}}					% transfer-function correlation
\safemath{\fcorr}{c_{H}}						% frequency-correlation function

\safemath{\mi}{I}							% mutual information
\safemath{\capacity}{C}						% capacity

\safemath{\normal}{\mathcal{N}}			% normal distribution
\safemath{\jpg}{\mathcal{CN}}			% jointly proper Gaussian
\safemath{\mchain}{\leftrightarrow}		% Markov chain
\safemath{\dB}{\,\mathrm{dB}}
\safemath{\dBm}{\,\mathrm{dBm}}
\safemath{\Hz}{\,\mathrm{Hz}}
\safemath{\kHz}{\,\mathrm{kHz}}
\safemath{\MHz}{\,\mathrm{MHz}}
\safemath{\GHz}{\,\mathrm{GHz}}
\safemath{\s}{\,\mathrm{s}}
\safemath{\ms}{\,\mathrm{ms}}
\safemath{\mus}{\,\mathrm{\text{\textmu}s}}
\safemath{\ns}{\,\mathrm{ns}}
\safemath{\ps}{\,\mathrm{ps}}
\safemath{\meter}{\,\mathrm{m}}
\safemath{\mm}{\,\mathrm{mm}}
\safemath{\cm}{\,\mathrm{cm}}
\safemath{\m}{\,\mathrm{m}}
\safemath{\W}{\,\mathrm{W}}
\safemath{\mW}{\, \mathrm{mW}}
\safemath{\J}{\,\mathrm{J}}
\safemath{\K}{\,\mathrm{K}}
\safemath{\bit}{\,\mathrm{bit}}
\safemath{\nat}{\,\mathrm{nat}}

\safemath{\define}{\triangleq}			% definition

\safemath{\equivalent}{\sim}
\safemath{\distas}{\sim}					% distributed according to
\safemath{\sdiff}{\Delta}				% symmetric set difference

\safemath{\reals}{\mathbb{R}}
\safemath{\positivereals}{\reals_{+}}
\safemath{\integers}{\mathbb{Z}}
\safemath{\posint}{\integers_{+}}
\safemath{\naturals}{\mathbb{N}}
\safemath{\posnaturals}{\naturals_{+}}
\safemath{\complexset}{\mathbb{C}}
\safemath{\rationals}{\mathbb{Q}}

\newcommand*{\fancyrefapplabelprefix}{app}		% Appendix
\newcommand*{\fancyrefthmlabelprefix}{thm}		% Theorem
\newcommand*{\fancyrefremlabelprefix}{rem}		% Remark
\newcommand*{\fancyreflemlabelprefix}{lem}		% Lemma
\newcommand*{\fancyrefcorlabelprefix}{cor}		% Corollary
\newcommand*{\fancyrefdeflabelprefix}{def}		% Definition
\newcommand*{\fancyrefproplabelprefix}{prop}		% Proposition
\newcommand*{\fancyrefexmpllabelprefix}{exmpl}
\newcommand*{\fancyrefalglabelprefix}{alg}		% Algorithm
\newcommand*{\fancyreftbllabelprefix}{tbl}		% Algorithm

\frefformat{vario}{\fancyrefseclabelprefix}{Sec.~#1}
\frefformat{vario}{\fancyrefthmlabelprefix}{Thm.~#1}
\frefformat{vario}{\fancyrefremlabelprefix}{Rem.~#1}
\frefformat{vario}{\fancyreftbllabelprefix}{Tbl.~#1}
\frefformat{vario}{\fancyreflemlabelprefix}{Lem.~#1}
\frefformat{vario}{\fancyrefcorlabelprefix}{Cor.~#1}
\frefformat{vario}{\fancyrefdeflabelprefix}{Def.~#1}
\frefformat{vario}{\fancyreffiglabelprefix}{Fig.~#1}
\frefformat{vario}{\fancyrefapplabelprefix}{App.~#1}
\frefformat{vario}{\fancyrefeqlabelprefix}{(#1)}
\frefformat{vario}{\fancyrefproplabelprefix}{Prop.~#1}
\frefformat{vario}{\fancyrefexmpllabelprefix}{Ex.~#1}
\frefformat{vario}{\fancyrefalglabelprefix}{Alg.~#1}

\newcommand{\marginnote}[1]{\marginpar{\framebox{\framebox{#1}}}}
\newcommand{\comment}[2]{[\marginnote{#1}\textit{#1}:\ \textit{#2}]}

 \newtheorem{thm}{Theorem}
 \newtheorem{rem}{Remark}
 \newtheorem{prop}{Proposition}

\safemath{\dictab}{[\,\dicta\,\,\dictb\,]}

\safemath{\ysig}{\bmy}
\safemath{\ysighat}{\hat{\ysig}}
\safemath{\ysigdim}{M}
\safemath{\xsig}{\bmx}
\safemath{\xsigdim}{N}
\safemath{\nx}{n_x}
\safemath{\zsig}{\bmz}
\safemath{\zsigdim}{\ysigdim}
\safemath{\rsig}{\bmr}
\safemath{\Adict}{\bA}
\safemath{\Adicttilde}{\widetilde{\Adict}}
\safemath{\Adictdim}{\outputdim\times\xsigdim}
\safemath{\avec}{\bma}
\safemath{\avectilde}{\tilde{\avec}}
\safemath{\Bdict}{\bB}
\safemath{\Bdicttilde}{\widetilde{\Bdict}}
\safemath{\Cdict}{\bC}
\safemath{\cvec}{\bmc}
\safemath{\Ddict}{\bD}
\safemath{\Ddictdim}{\ysigdim\times\xsigdim}
\safemath{\dvec}{\bmd}
\safemath{\Ddicttilde}{\widetilde{\bD}}
\safemath{\Bonb}{\bB}
\safemath{\bvec}{\bmb}
\safemath{\Bonbdim}{\ysigdim\times\ysigdim}
\safemath{\noise}{\bmn}
\safemath{\noisedim}{\ysigim}
\safemath{\err}{\bme}
\safemath{\errdim}{\ysigdim}
\safemath{\errset}{\setE}
\safemath{\nerr}{n_e}
\safemath{\delop}{\bP_\errset}
\safemath{\delopc}{\bP_{{\errset}^c}}

\safemath{\cplxi}{\imath}
\safemath{\cplxj}{\jmath}
\safemath{\dict}{\matD}
\safemath{\inputdim}{N}		% number of columns of dictionary D
\safemath{\outputdim}{M}		%number of rows of dictionary D
\safemath{\sparsity}{S}	%sparsity
\safemath{\inputdimA}{{N_a}}	%total number of elements in dictionary A
\safemath{\inputdimB}{{N_b}}	%total number of elements in dictionary B
\safemath{\elemA}{{n_a}}	%number of elements chosen from dictionary A
\safemath{\elemB}{{n_b}}	%number of elements chosen from dictionary B
\safemath{\resA}{\matR_a}	%restriction map to elements of dictionary A
\safemath{\resB}{\matR_b}	%restriction map to elements of dictionary B
\safemath{\subD}{\matS} %subdictionary
\safemath{\subA}{\matS_a} %subdictionary part of A
\safemath{\subB}{\matS_b} %subdictionary part of B
\safemath{\dicta}{\matA} 	% first subdictionary
\safemath{\dictb}{\matB} 	% second subdictionary
\safemath{\hollowS}{H}
\safemath{\hollowA}{H_a}
\safemath{\hollowB}{H_b}
\safemath{\cross}{Z}
\safemath{\coh}{\mu_d}			% coherence dictionary
\safemath{\coha}{\mu_a}			% coherence first subdictionary
\safemath{\cohb}{\mu_b}			% coherence second subdictionary
\safemath{\mubs}{\nu}	%block sub-coherence
\safemath{\cohm}{\mu_m} %mutual coherence
\safemath{\dictset}{\setD}	% set of dictionaries
\safemath{\dictsetp}{\dictset(\coh,\coha,\cohb)}	% set of dictionaries parametrized
\safemath{\dictsetgen}{\dictset_\text{gen}}
\safemath{\dictsetgenp}{\dictsetgen(\coh)}
\safemath{\dictsetonb}{\dictset_\text{onb}}
\safemath{\dictsetonbp}{\dictsetonb(\coh)}

\safemath{\leftside}{U}
\safemath{\rightsideA}{R_a}
\safemath{\rightsideB}{R_b}

\safemath{\indexS}{\setI_S} %set of indices participating in sub-dictionary S

\safemath{\na}{n_a}			% cardinality of set of linearly independent columns of first dictionary
\safemath{\nb}{n_b}			% cardinality of set of linearly independent columns of second dictionary
\safemath{\coeffa}{p_i}	%coefficients for columns of A
\safemath{\coeffb}{q_j}	%coefficients for columns of B
\safemath{\seta}{\setP}		% set of linearly independent columns of A
\safemath{\setb}{\setQ}     % set of linearly independent columns of B
\safemath{\setw}{\setW}	%set of n largest elements of w
\safemath{\setz}{\setZ}	%set of L-n largest elements of z
\safemath{\cola}{\veca}		% generic element of the dictionary A
\safemath{\colb}{\vecb}		% generic element of the dictionary B
\safemath{\cold}{\vecd}		% generic element of the dictionary D
\safemath{\inputvec}{\vecx} 	%coefficient vector (input)
\safemath{\error}{\vece}	%error vector
\safemath{\noiseout}{\vecz} 	%noisy output vector
\safemath{\inputvecel}{x}
\safemath{\inputveca}{\vecx_a}
\safemath{\inputvecb}{\vecx_b}
\safemath{\outputvec}{\vecy}	%output of Dictionary
\safemath{\lambdamin}{\lambda_{\mathrm{min}}}
\safemath{\elltwo}{\ell_2}
\safemath{\ellone}{\ell_1}
\safemath{\ellzero}{\ell_0}
\safemath{\ellinf}{\ell_\infty}
\safemath{\ellinftilde}{\ell_{\widetilde\infty}}
\safemath{\licard}{Z(\coh,\coha,\cohb)}
\safemath{\xsol}{\hat{x}}
\safemath{\xbord}{x_b}		%Solution at the border
\safemath{\xstat}{x_s}		%Solution stationary in l0 prob
\safemath{\xstatLone}{\tilde{x}_s}
\safemath{\order}{\mathcal{O}} %order notation (big O)
\safemath{\scales}{\Theta} %scales as
\safemath{\ones}{\mathbf{1}} %all ones matrix
\safemath{\zeroes}{\mathbf{0}} %all zeroes matrix
\safemath{\thlone}{\kappa(\coh,\cohb)} %treshold l1 problem
\safemath{\constoneA}{\delta} %constant in l1 theorem to save space
\safemath{\constoneB}{\epsilon} %constant in l1 theorem to save space
\safemath{\nlarge}{L}				   %num large elements
\safemath{\sumlarge}{S_\nlarge}
\safemath{\maxlarger}{P_\nlarge}	   % maximum in Gribonval and Nielsen
\safemath{\Pzero}{\textrm{P0}}	
\safemath{\Pone}{\textrm{P1}}
\safemath{\vecfir}{\vecw}			 % \vecv element of the kernel of the dictionary, \vecv=[\vecfir \vecsec]
\safemath{\vecsec}{\vecz}
\safemath{\elvecfir}{w}              % element of vecfir
\safemath{\elvecsec}{z}				 % element of vecsec
\safemath{\nlargefir}{n}
\safemath{\normout}{\gamma}
\safemath{\auxfun}{h}
\safemath{\supp}{\textrm{supp}}%support

\safemath{\indexa}{\ell}
\safemath{\indexb}{r}
\safemath{\indexc}{i}
\safemath{\indexd}{j}

\safemath{\project}{P}%projector

\usepackage{framed}

\renewcommand{\bSigma}{\mathbf{\Sigma}}

\renewcommand{\bLambda}{\mathbf{\Lambda}}

\safemath{\sfp}{\textsf{p}}
\safemath{\sfc}{\textsf{c}}

\IEEEoverridecommandlockouts
\allowdisplaybreaks % THIS ALLOWS EQUATIONS TO BREAK THE PAGE

\newif\ifshowcomments

\showcommentstrue

\ifshowcomments

\newcommand{\gm}[1]{\comment{gm}{\textcolor{orange}{#1}}}
\newcommand{\cs}[1]{\comment{cs}{\textcolor{red}{#1}}}
\newcommand{\as}[1]{\comment{as}{\textcolor{blue}{#1}}}

\else

\newcommand{\gm}[1]{}
\newcommand{\cs}[1]{}
\newcommand{\as}[1]{}

\fi

\safemath{\compquant}{\mathcal{Q}}

\begin{document}
\bstctlcite{IEEEexample:BSTcontrol} % keep repeated author names in the biography

\title{Fundamental Limits for Jammer-Resilient Communication in Finite-Resolution MIMO}

\author{
	\IEEEauthorblockN{Gian Marti, Alexander Stutz-Tirri, and Christoph Studer}
 	\IEEEauthorblockA{\em Department of Information Technology and Electrical Engineering, ETH Zurich, Switzerland\\
              email: gimarti@ethz.ch, alstutz@iis.ee.ethz.ch, and studer@ethz.ch}	
}

\maketitle

\begin{abstract}
Spatial filtering based on multiple-input multiple-output (MIMO) processing is a powerful method for jammer mitigation.
In principle, a MIMO receiver can null the interference of a single-antenna jammer at the cost of only one degree of freedom---if the number of receive antennas is large, communication performance is barely affected. 
In this paper, we show that the potential for MIMO jammer mitigation based on the digital outputs of finite-resolution 
analog-to-digital converters (ADCs) is fundamentally worse: Strong jammers will either cause the ADCs to saturate
(when the ADCs' quantization range is small) or drown legitimate communication signals in quantization noise 
(when the ADCs' quantization range is large).
We provide a fundamental bound on the mutual information between the quantized receive signal and the legitimate transmit 
signal. Our bound shows that, for any fixed ADC resolution,  
the mutual information tends to zero as the jammer power tends to infinity, \emph{regardless of the quantization strategy}. 
Our bound also confirms the intuition that for every 6.02\,dB increase in jamming power, the ADC resolution must be increased by 1\,bit in order to prevent the mutual information from vanishing.
\end{abstract}

\vspace{-0.5mm}
\section{Introduction}\vspace{-0.5mm}
Jammers are natural enemies of wireless communication systems \cite{pirayesh2022jamming}. 
A powerful countermeasure is spatial filtering via multiple-input multiple-output (MIMO) processing\mbox{\cite{leost2012interference, marti2023universal}.}
If the receive signal at a $B$-antenna receiver consists of the linear superposition of 
legitimate communication signals and low-rank jammer interference, then the receiver
can null the jammer interference at the cost of $I$ degrees of freedom (where $I$ is 
the rank of the jammer interference). If $B\gg I$ and the wireless channel is well conditioned, 
then communication performance is barely affected by the jammer. 
However, if the receive signal is converted to digital using 
analog-to-digital converters (ADCs) before equalization, then the relation between 
the transmit signals and the finite-resolution digital receive signals is no longer linear \cite{mollen16c, jacobsson17b, demir2020bussgang}. 
In particular, strong jammers cause the ADCs to saturate when the ADCs' quantization 
range is small, and they drown legitimate communication signals in quantization noise 
\mbox{when the ADCs' quantization range is large \cite{block2006performance}.}

\vspace{-0.5mm}
\subsection{Contributions}\vspace{-0.5mm}
We show that quantization \emph{fundamentally} limits jammer-resilient communication in MIMO systems. To this end, we derive a bound on the mutual information between the quantized receive 
signal and the legitimate transmit signal as a function of the ADC resolution and the receive 
signal-to-interference-plus-noise ratio (SINR). Our bound holds for \emph{any} 
ADCs with a given resolution (not only for uniform quantizers, 
and regardless of what kind of automatic gain control is applied) 
and for \emph{any} power-limited signaling at the legitimate transmitters (not only for QAM or Gaussian signals).
Our bound reveals that, for any fixed ADC resolution, the mutual information tends to zero as the jammer power 
tends to infinity. Furthermore, our results rigorously confirm the intuition that 
the ADC resolution must be increased by 1\,bit for every 6.02\,dB increase in jamming power 
to prevent the mutual information from vanishing.
We complete our analysis with numerical simulations and with a comparison to the behavior of infinite-resolution 
MIMO communication systems (which are analyzed in \fref{app:unquantized}).

\vspace{-0.5mm}
\subsection{Notation}\vspace{-0.5mm}
Matrices, vectors, and scalars are denoted by boldface uppercase (e.g., $\bA$),
boldface lowercase (e.g., $\bma$), and italic or sans-serif 
(e.g., $a$, $A$, and $\mathsf{A}$), respectively. 
For a matrix~$\bA$, the transpose is $\tp{\bA}$, the (entry-wise) conjugate is $\bA^{\ast}$,
the conjugate tranpose is~$\herm{\bA}$, the determinant is~$|\bA|$, the space spanned by its columns
is $\textit{col}(\bA)$, and the $n$th row is $\bma_{(n)}$. 
The Euclidean norm is $\|\cdot\|_2$. 
A circularly-symmetric complex Gaussian vector with covariance matrix~$\bC$ is 
denoted $\bma\sim\setC\setN(\mathbf{0},\bC)$. Whether a quantity is a random variable
or a constant becomes clear from the context. 
Mutual information, conditional mutual information, discrete entropy, conditional discrete entropy, 
differential entropy, and conditional differential entropy are
denoted by $I(\cdot\,;\cdot)$, $I(\cdot\,;\cdot\,|\,\cdot)$, $H(\cdot)$, $H(\cdot|\cdot)$, $h(\cdot)$, and $h(\cdot\,|\,\cdot)$, 
respectively. 
Logarithms are to the basis $2$. 
The binary entropy function is $H_b(x)=-x\log(x)-(1\!-\!x)\log(1\!-\!x)$ and we use
$\bar{H}_b(x) = H_b(\min\{x,\frac12\})$.
The (complementary) error function is $\erf(x)=2\pi^{-\sfrac{1}{2}}\int_{0}^{x}\exp(-t^2)\,\textnormal{d}t$ and 
$\erfc(x)=1-\erf(x)$, respectively.
$[N]$ is the set of integers from $1$ through $N$ and $\lfloor x \rfloor$ is the largest integer no greater than $x$. 
$\mathbbm{1}\{\textrm{statement}\}$ equals~$1$ if statement is true and $0$ otherwise.
For the functions $f$ and $g$, the relation $f\prec g$ means that $\exists \epsilon>0:\lim_{x\to\infty}\frac{f(x)}{g(x)} x^{\epsilon}=0$, and $f\succeq g$ is its logical negation. The relation $f\prec_{\scriptscriptstyle{+}}g$
means that $\exists \epsilon>0: \lim_{x\to\infty}(1+\epsilon)f(x)-g(x)=-\infty$, and $f\succeq_{\scriptscriptstyle{+}}g$
is its logical negation.
We use $\inf_{\setA}f$ as shorthand for $\inf_{x\in\setA}f(x)$. 

\vspace{-0.5mm}
\section{System Model} \label{sec:model}\vspace{-0.5mm}
We consider a frequency-flat MIMO system under jamming. 
We focus on a multi-user (MU) MIMO uplink scenario, but our results are easily extended to 
single-input multiple-output or point-to-point MIMO scenarios. 
We model the time-sampled receive signal $\bmy\in\opC^B$ at a $B$-antenna basestation (BS)~as
\begin{align}
	\bmy &= \bH\bms + \bJ\bmw + \bmn, \label{eq:io1}
\end{align}
where $\bH \in \opC^{B\times U}$ is the channel matrix of $U$ legitimate single-antenna user equipments (UEs) 
with data signal $\bms\in\opC^U$,
$\bJ \in\opC^{B\times I}$ is the channel matrix of an \mbox{$I$-antenna} jammer with transmit vector \mbox{$\bmw_k\in\opC^{I}$},
and $\bmn\sim\setC\setN(\mathbf{0},\No\bI_B)$ is circularly- symmetric complex white Gaussian noise.
We assume that the matrices $\bH$ and $\bJ$ stay constant over an extended time period (slow fading) 
and are perfectly known at the receiver.\footnote{The assumption of perfect channel knowledge strengthens
our result: An upper bound on possible rates of communication~with perfect channel knowledge also holds when the channels are not known, or known imperfectly.}
The random vectors $\bms$, $\bmw$, and $\bmn$ are assumed to be mutually independent. 
By defining the receive data signal $\bmr\triangleq\bH\bms$ and the receive interference $\bmz\triangleq\bJ\bmw$, 
we can rewrite \eqref{eq:io1} as
\begin{align}
	\bmy &= \bmr + \bmz + \bmn. \label{eq:io2}
\intertext{
We then model analog-to-digital conversion at the BS by quantizing the time-sampled receive signal $\bmy$ in amplitude~using
}
	\bmq &= \compquant(\bmy), \label{eq:vec_quant}
\end{align}
where 
\begin{align}
	&\compquant: \opC^B \to \mathfrak{Q}^{2B}, 
	~\bmy \mapsto \big(q_{(1,\mathfrak{r})},q_{(1,\mathfrak{i})}, \dots, q_{(B,\mathfrak{r})}, q_{(B,\mathfrak{i})}\big)
\end{align}
% with 
\begin{align}
    q_{c} = Q_c(y_c), \quad c\in\setC\triangleq [B]\times\{\mathfrak{r},\mathfrak{i}\},
\end{align}
% is a quantization function that 
quantizes the in-phase and quadrature parts (indexed by $\mathfrak{r}$ and~$\mathfrak{i}$, respectively) of
the receive signal $\bmy$ using scalar quantizers $Q_c: \mathbb{R} \to \mathfrak{Q}_c$ with \mbox{$M\triangleq|\mathfrak{Q}_c|\geq2$}
quantization levels. The scalar quantizers~$Q_c$ may be non-uniform and different for every $c\in\setC$, 
but we assume that the ADC resolution $M$ is the same for all $c\in\setC$.
Equations~\eqref{eq:io2} and~\eqref{eq:vec_quant} can be restated for the scalar signal components $c\in\setC$ as
\begin{align}
	y_c &= r_c + z_c + n_c \label{eq:comp_io1} \\
	q_c &= Q_c(y_c). \label{eq:comp_io2}
\end{align}

\section{Main Result}

We state our main result in three steps; all proofs are in~\fref{app:proofs}. 
First, we relate the mutual information between $\bmq$ and $\bms$ 
to the sum over conditional mutual informations between the signal components $r_c$ and the ADC outputs $q_c$, 
given the interference-plus-noise components $z_c+n_c$:

\begin{thm} \label{thm:mutual_inf}
Consider the model from \fref{sec:model}. Then
\begin{align}
	\!\!\!I(\bmq; \bms) 
	&\!\leq\! \min\!\Big\{\! \sum_{c\in\setC}  I \big(q_c; r_c| z_c \!\!+\! n_c\big)
	, \log \!\big| \No^{-1}\bH\herm{\bH}\!\!\!+\!\bI_B\big|\Big\} .\!\! \label{eq:thm1}
\end{align}	
\end{thm}
(The second argument in the minimum comes from the fact that the mutual information is always 
bounded by the capacity of an unquantized, jammer-free system; cf. \fref{app:unquantized}.)
Second, we bound the conditional mutual information in~\eqref{eq:thm1} in terms of
the number of quantization levels and the SINR:

\begin{thm} \label{thm:quant}
Consider the model from \fref{sec:model} for
$z_c\!\sim\setN(0,\mathsf{Z}_c)$, $n_c\!\sim\setN(0,\No)$, and 
$\Ex{}{|r_c|^2} \!\leq \mathsf{R}_c$. 
Let 
\begin{align} \label{eq:sinr}
\textnormal{\textsf{SINR}}_c\triangleq \frac{\mathsf{R}_c}{\mathsf{Z}_c+\No}
\end{align}
denote the SINR at the $c$th ADC and define 
\begin{align} 
	\bar{f}(M,\mathsf{SINR}_c) &= \erf\!\bigg(\! \frac{(M\!-\!1)\sqrt{\mathsf{SINR}_c}}{\sqrt{2}} \bigg) \nonumber\\
	&\hspace{-4.9mm} + \sqrt{\frac{2}{\pi}}(M\!-\!1)\sqrt{\mathsf{SINR}_c} \exp\!\bigg(\!\!-\frac{(M\!-\!1)^2\mathsf{SINR}_c}{2} \!\bigg) \nonumber\\
	&\hspace{-4.9mm} - (M\!-\!1)^2\, \mathsf{SINR}_c \erfc\!\bigg(\! \frac{(M\!-\!1)\sqrt{\mathsf{SINR}_c}}{\sqrt{2}} \!\bigg)
	\label{eq:f_bound}
\end{align}
and 
\begin{align} 
	 \hspace{-2mm}\bar{\iota}(M,\mathsf{SINR}_c)&\triangleq \min\!\Big\{\! \log M, \nonumber\\
    \bar{H}_b\big(&\bar{f}(M, \mathsf{SINR}_c)\big) \!+\! \bar{f}(M,\mathsf{SINR}_c)\log(M\!-\!1) \!\Big\}.\!\! 
    \label{eq:cond_entropy_bound}
\end{align}
Then, the conditional mutual information $ I(q_c, r_c| z_c + n_c )$ is bounded by
\begin{align}
    I(q_c; r_c| z_c + n_c ) \leq \bar{\iota}(M,\mathsf{SINR}_c).
\end{align}
\end{thm}

\begin{rem} \label{rem:simplified}
The statement of \fref{thm:quant} also holds when, in \eqref{eq:cond_entropy_bound},  
$\bar{f}(M,\mathsf{SINR}_c)$ is replaced with its upper bound
\begin{align}
  \bar{\bar{f}}(M,\mathsf{SINR}_c) = \sqrt{\frac{8}{\pi}} \, (M\!-\!1) \sqrt{\mathsf{SINR}_c}.
\end{align}
A derivation of this upper bound is shown in \fref{app:proof_6db}.
\end{rem}

The final step combines the results from \fref{thm:mutual_inf} and \fref{thm:quant}: 

\begin{thm} \label{thm:main}
Consider the model from \fref{sec:model}, and assume that $\bmw\sim\setC\setN(\mathbf{0},\rho\bI_I)$, that $\bmn\sim\setC\setN(\mathbf{0},\No\bI_B)$, 
and that the entries $s_u$ of $\bms$ are uncorrelated and satisfy $\Ex{}{|s_u|^2} \leq 1$. Then 
%\blue{$I(\bmq; \bms)\leq\min\{I(\bmq; \bms), \log \big| \No^{-1}\bH\herm{\bH}+\bI_B\big|\}$, where
\begin{align}
	 \!\!\!I(\bmq; \bms) &\leq\bar{I}(M,\rho,\bH,\bJ) \label{eq:mutual_inf_mainbound} \\
& \triangleq \min\!\Big\{\!\sum_{c\in\setC}  \bar{\iota}(M, \overline{\mathsf{SINR}}_c), \log \big| \No^{-1}\bH\herm{\bH}\!\!+\!\bI_B\big|\!\Big\},\!  \label{eq:mi_bound_case_distinction}
\end{align}
where $\overline{\mathsf{SINR}}_c\geq\mathsf{SINR}_c$ is an upper bound on the SINR at the $c$th ADC which, 
for $c\in\{(b,\mathfrak{r}),\,(b,\mathfrak{i})\}$, is given by
\begin{align}
	\overline{\mathsf{SINR}}_c = 
		\frac{2\|\bmh_{(b)}\|_2^2}{\rho\|\bmj_{(b)}\|_2^2\!+\!\No}.
	\label{eq:sinr_bound}
\end{align}
\end{thm}

\begin{rem} \label{rem:sinr}
If $\bms\sim\setC\setN(\mathbf{0},\bI_U)$, then \eqref{eq:sinr_bound} can be tightened to 
\begin{align}
	\mathsf{SINR}_c = \overline{\mathsf{SINR}}_c = 
		\frac{\|\bmh_{(u)}\|_2^2}{\rho\|\bmj_{(u)}\|_2^2\!+\!\No}.
\end{align}
\end{rem}

\section{Analysis} \label{sec:analysis}
\subsection{Asymptotic Analysis} \label{sec:asymptotic}
For fixed channel realizations $\bH$ and $\bJ$, and for a fixed thermal noise power $\No$,  
the bound in \eqref{eq:mutual_inf_mainbound} is a function of the jammer power $\rho$ 
(via \eqref{eq:sinr_bound}) and the number of ADC quantization levels $M$. 
Two questions are therefore natural to ask:\footnote{The attentive reader will notice that the 
second question presupposes the answer to the first question.}
\begin{enumerate}
	\item What happens in the limit $\rho\to\infty$ when the ADC resolution $M$ is fixed?
	\item How does the ADC resolution $M$ have to scale with $\rho$ to prevent the mutual information $I(\bmq; \bms)$
		from vanishing?
\end{enumerate}
The answers to these questions are as follows:

\begin{prop} \label{prop:vanish}
Consider the model from \fref{sec:model}, and assume 
that $\bmw\sim\setC\setN(\mathbf{0},\rho\bI_I)$, $\bmn\sim\setC\setN(\mathbf{0},\No\bI_B)$, 
and that the entries $s_u$ of $\bms$ are uncorrelated and satisfy $\Ex{}{|s_u|^2} \leq 1$. 
If all rows of $\bJ$ are nonzero, then 
$I(\bmq; \bms) \xrightarrow{\rho\to\infty} 0$.
\end{prop}

\begin{prop} \label{prop:6dB}
Consider the model from \fref{sec:model}, and assume 
that $\bmw\sim\setC\setN(\mathbf{0},\rho\bI_I)$, that $\bmn\sim\setC\setN(\mathbf{0},\No\bI_B)$, 
and that the entries $s_u$ of $\bms$ are uncorrelated and satisfy $\Ex{}{|s_u|^2} \leq 1$. 
If~all rows of $\bJ$ are nonzero, then $\lim_{\rho\to\infty}I(\bmq; \bms)>0$ implies
\begin{align}
	M \succeq \sqrt{\rho}. \label{eq:asymp_rho}
\end{align}
Equation \eqref{eq:asymp_rho} can be restated in the logarithmic domain as 
\begin{align}
  \log M \succeq_{\scriptscriptstyle{+}} \frac{\log10}{20} \rho_{\textnormal{dB}} \approx \frac{1}{6.02}\rho_{\textnormal{dB}},
\end{align}
which implies that one additional bit of ADC resolution is needed for every $6.02$\,dB increase in jammer power! 
\end{prop}

\subsection{Numerical Analysis}\label{sec:numeric}

\begin{figure}[tp]
	\centering
	\includegraphics[width=1.01\columnwidth]{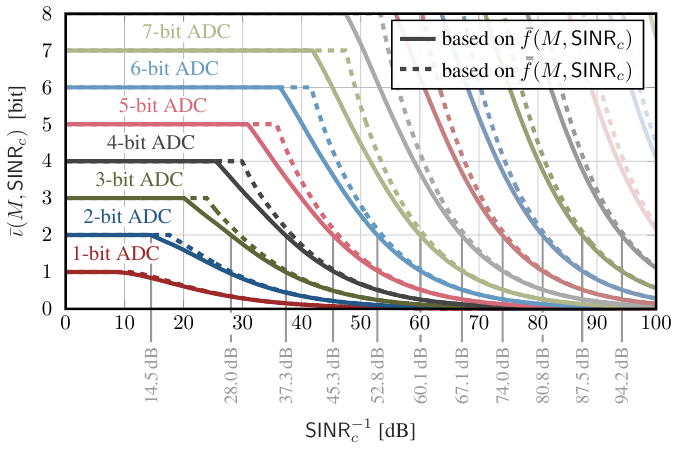}
 \vspace{-0.6cm}
	\caption{The bound from \fref{thm:quant} as a function of the reciprocal signal-to-interference-plus-noise ratio 
    $\inv{\mathsf{SINR}_c}$. The number $M$ of ADC quantization levels is expressed by the number of equivalent bits $\log M$. 
    Dashed curves are based on the weaker (but more intuitive) bound from \fref{rem:simplified}.
    }
    \vspace{-2mm}
	\label{fig:adc_entropy}
\end{figure}

We turn to the numerical analysis of our results. 
We start by plotting the bound $\bar{\iota}(M,\mathsf{SINR}_c)$ from \fref{thm:quant} as a function of 
the reciprocal signal-to-interference-plus-noise ratio~$\mathsf{SINR}_c^{-1}$, for different numbers of quantization 
levels $M$ (note that the number of quantization levels for an $m$-bit quantizer is \mbox{$M=2^m$}).
\fref{fig:adc_entropy} depicts the bounds $\bar{\iota}(M,\mathsf{SINR}_c)$ from \fref{thm:quant} (solid~lines)
as well as the (slightly) looser bounds based on the easier-to-interpret function $\bar{\bar{f}}(M,\mathsf{SINR}_c)$ 
from \fref{rem:simplified} (dashed lines). 
For any fixed ADC resolution, \fref{fig:adc_entropy} shows the following: The conditional (on the interference plus noise)
entropy of an ADC output is guaranteed to be strictly smaller than its number of output bits when the 
$\mathsf{SINR}_c^{-1}$ at that ADC's 
input exceeds a certain value; and the conditional mutual information tends to zero with increasing $\mathsf{SINR}_c^{-1}$.
An ADC with a large $\mathsf{SINR}_c^{-1}$ at its input will therefore barely contribute to the mutual information $I(\bmq;\bms)$; cf. \fref{thm:main}.

\fref{fig:adc_entropy} also exemplarily marks (in gray) the $\mathsf{SINR}_c^{-1}$ levels at which the conditional 
mutual information of an ADC falls below 2~bits: If $\mathsf{SINR}_c^{-1}$ exceeds 74.0\,dB, then the conditional entropy of
a 9-bit quantizer is smaller than 2 bits; if $\mathsf{SINR}_c^{-1}$ exceeds 80.8\,dB, then the conditional entropy of
a 10-bit quantizer is smaller than 2 bits. These results point towards the result from \fref{prop:6dB}: 
Asymptotically, the ADC resolution must increase by one bit for every 6.02\,dB of increased jamming power 
to prevent its conditional mutual information (and thus the overall mutual information; cf.~\fref{thm:mutual_inf}) from~vanishing.

%\vspace{2mm}

Next, we evaluate the bound on $I(\bmq;\bms)$ from \fref{thm:main} as a function of the channel realizations~$\bH$ and~$\bJ$. 
To this end, we consider a MU-MIMO system with $B=16$ receive antennas (corresponding to $|\setC|=32$ receive ADCs), 
$U=2$ UEs, and an $I=1$ antenna jammer that transmits Gaussian symbols with $\rho=60$\,dB more power than each of the UEs. 
The signal-to-noise ratio (SNR) is assumed to be~$30$\,dB (i.e., we have $\No=-30$\,dB, meaning that the noise on its own plays 
no significant role).
We consider two different channel models: (i)~Rayleigh fading, where the entries of $\bH$ and $\bJ$ are drawn i.i.d. 
according to $\setC\setN(0,1)$; and (ii) a textbook line-of-sight (LoS) model in which the BS antennas are arranged in a 
uniform linear array (ULA) with half-wavelength antenna spacing and where the UEs and the jammer are placed uniformly 
independently in a $120^{\circ}$ sector in front of the BS, so that the channel vector between the BS and a device placed 
at angle $\theta$ is
\begin{align}
	\tp{\big[1, e^{-i\pi\cos(\theta)}, \dots, e^{-i\pi\cos(\theta)(B-1)} \big]}. \label{eq:los}
\end{align}
Note that we do not model path loss explicitly in either the Rayleigh fading or the LoS model: Differences in 
path loss between the UEs are assumed to be compensated by UE power control, and differences in path loss between 
the UEs and the jammer are absorbed into the jammer transmit power $\rho$.

\begin{figure}[tp]
	\centering
    \vspace{-1mm}
    \subfigure[Rayleigh]{
        \hspace{-5mm}
        \includegraphics[height=3.8cm]{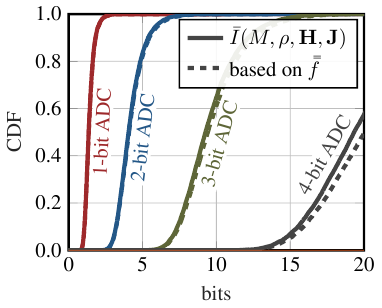}\label{fig:rayleigh}
        }
	\hspace{-3mm}
	\subfigure[LoS]{\
        \includegraphics[height=3.8cm]{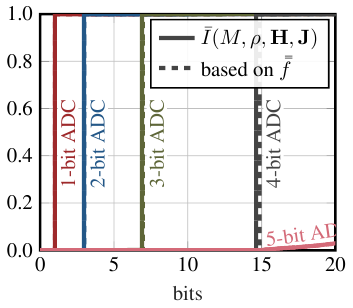}\label{fig:los}
            \hspace{-5mm}
            }
    \vspace{-1mm}
    \caption{Cumulative distribution functions (CDFs) over channel realizations of the bound from \fref{thm:main}, 
    for two different channel models.
    The number $M$ of ADC quantization levels is expressed by the number of equivalent bits $\log M$. 
    Dashed curves are based on the weaker bound from \fref{rem:simplified}.
    }	\label{fig:cdfs}
\end{figure}

In \fref{fig:cdfs}, we plot the cumulative distribution function (CDF) of $\bar{I}(M,\rho,\bH,\bJ)$ over random channel realizations.
\fref{fig:rayleigh} shows that, for $90\%$ of Rayleigh fading channel realizations, the mutual information 
$I(\bmq;\bms)$ under 1-bit quantization is lower than $2$ bits. In these cases, the jammer can reduce the mutual information 
between the quantized receive signal and the data signal by more than $\sfrac{30}{32}=93.75\%$ (because the $|\setC|=32$\, 
1-bit ADCs in principle have the potential to deliver $32$ bits of information about $\bms$).\footnote{In this argument, 
we assume that the thermal noise plays essentially no role here in reducing the mutual information.
This assumption is validated by the fact that the CDFs from \fref{fig:cdfs} are virtually identical when the noise~$\bmn$ is set to zero---the noise $\bmn$ contributes virtually nothing to the SINRs underlying the CDFs from \fref{fig:cdfs} since
it has orders of magnitude less energy than $\bmw$.
}
Hence, for these channel realizations, the possible rates of communication are reduced by more than $93.75\%$ compared to 
a jammer-free scenario \emph{regardless} of any potential jammer-mitigating equalization, linear or non-linear, that is 
applied to the digital signal $\bmq$. 
This behavior contrasts sharply with the jammer resilience of infinite-resolution MIMO communication systems with $\bmq=\bmy$, 
for which lower bounds on $I(\bmy; \bms)$ can be obtained that are independent of the jammer power, 
and that are close to the mutual information of equivalent jammer-free systems; see \fref{app:unquantized} for the details.

\fref{fig:los} shows results for LoS channels, which are qualitatively similar the results of \fref{fig:rayleigh}. 
However, for the textbook LoS model in \eqref{eq:los}, the norms of rows of the channel matrices~$\bH$ and $\bJ$
are deterministic, $\|\bmh_{(b)}\|_2^2=U$, and $\|\bmj_{(b)}\|_2^2=I$. 
As a result, $\bar{I}(M,\rho,\bH,\bJ)$ in \eqref{eq:mutual_inf_mainbound} becomes deterministic
and the CDFs turn into step-functions that depend only on $M$ and $\rho$
(provided that the second argument of the minimum in \eqref{eq:mi_bound_case_distinction} is not smaller than the first one; see the 5-bit ADC in \fref{fig:los}). 
\fref{fig:los} shows that the mutual information $I(\bmq;\bms)$ is lower than $1$ bit when 1-bit quantizers are used,
and lower than $3$ bits when 2-bit quantizers are used.

\section{Related Work}

Jamming attacks on MIMO communication systems and finite- or low-resolution
MIMO systems have independently been studied, e.g., in \cite{pirayesh2022jamming, marti2023universal, do18a} and 
\cite{mollen16c, jacobsson17b, li17b} as well as the references therein, but the 
interplay of these two phenomena has received comparably less attention.  
The first work that studies the interplay between jamming and finite-resolution quantization is \cite{block2006performance},
which analyzes the impact of out-of-band jammers on single-input single-output (SISO) communication systems with uniform quantizers. 
For such systems, and based on heuristic arguments, \cite{block2006performance} 
suggests a similar criterion to the one in \fref{prop:6dB} for successful communication.
The fact that jammer-resilient data detectors require higher resolution than non-resilient ones has also 
been observed empirically in \cite{bucheli2024vlsi}.
Reference \cite{lu2016effect} considers the effect of ADC quantization noise
due to jamming in the context of global navigation satellite systems (GNSSs) and suggests oversampling as~a~remedy.

More recently, a number of papers have considered the adverse impact of strong jamming on low-resolution MIMO communication systems and proposed various ways 
of removing a significant part of the jammer interference \emph{before} it is quantized by the ADCs \cite{pirzadeh2019mitigation, marti2021hybrid, marti2021snips, jiang2023ris}. 
Reference \cite{pirzadeh2019mitigation} proposes the use of spatial sigma-delta converters to form a spatial filter that removes the jammer interference in the analog domain. 
Reference \cite{marti2021hybrid} suggests to use a hybrid scheme consisting of an adaptive analog transform, which removes most of the jammer energy before it reaches the ADCs, and a subsequent digital equalizer that suppresses potential residues. 
For communication at millimeter-wave (mmWave) frequencies, \cite{marti2021snips} proposes a non-adaptive analog transform that sparsifies mmWave wireless channels and thus focuses the jammer interference on a subset of ADCs, so that the data signals can be reconstructed based on the outputs of the interference-free ADCs.
Reference \cite{jiang2023ris} suggests the use of a reconfigurable intelligent surface (RIS) to steer jamming signals away from the ADCs.  

Finally, \cite{teeti2021one} proposes a method for detecting jammers in MIMO systems with 1-bit quantizers, but does not consider methods
for their subsequent mitigation.  

% None of these mentioned works study the fundamental information-theoretic limits of low- or finite-resolution systems under jamming attacks. 
% \gm{maybe don't need to say this} 

\section{Conclusions}

We have derived fundamental information-theoretic limits for communication in finite-resolution MIMO under jamming. 
Our results have shown that, for any fixed ADC resolution, the mutual information between the legitimate communication signals and
the quantized receive signal tends to zero as the jammer power tends to infinity. 
This shows that finite-resolution MIMO systems are fundamentally vulnerable to jamming attacks in a way that infinite-resolution systems are~not. 
In consequence, it is beneficial to mitigate jammers already \emph{before} data conversion, e.g., using the methods proposed 
in~\cite{marti2021hybrid,marti2021snips}.

\appendices

\section{Results on the Jammer-Resilience of Infinite-Resolution MIMO} \label{app:unquantized}
\fref{prop:vanish} from \fref{sec:analysis} demonstrates that a single-antenna jammer can drive the rates at 
which data transmission to a low-resolution multi-antenna BS is possible to zero, 
regardless of any jammer-mitigating equalization that might be applied to the digital receive signal.
Moreover, the results from \fref{sec:numeric} show that a $60$\,dB single-antenna jammer 
can reduce the rates at which communication from $U=2$ UEs to a $B=16$ antenna BS 
with 1-bit ADCs is possible by more than $93.75\%$ compared to a jammer-free~scenario.

We now show that \emph{infinite-resolution} MIMO communication systems do not share such a 
vulnerability against single- or few-antenna jammers, provided that the spatial diversity is sufficient. 
For this, we consider a communication system as in \fref{sec:model}---except that, since the receiver is 
assumed to be infinite-resolution, we focus on the mutual information $I(\bmy;\bms)$ between $\bms$ and the \emph{unquantized}
receive signal $\bmy$. 
Our analysis will make use of the projection onto the orthogonal complement of the jammer subspace, 
given by the matrix $\herm{\bU_{\orth}}$, whose rows  form an orthonormal basis of the 
orthogonal complement of $\textit{col}(\bJ)$.\footnote{
Strictly speaking, $\herm{\bU_{\orth}}$ is not a projection---only $\bU_{\orth}\herm{\bU_{\orth}}$ is. 
We refer to $\herm{\bU_{\orth}}$ as a projection nonetheless.
}
This operator can be used as a spatial filter at the BS and has the property that $\herm{\bU_{\orth}}\bJ=\mathbf{0}$.
Since all channels are known, the BS knows $\herm{\bU_{\orth}}$.
We then have the following~results:

\begin{prop} \label{prop:unquant_mi}
Consider the input-output relation in \eqref{eq:io1}, and~assume that 
\mbox{$\bms\sim\setC\setN(\mathbf{0},\bI_U)$} and $\bmn\sim\setC\setN(\mathbf{0},\No\bI_B)$.
The distribution of the jammer signal $\bmw$ can be arbitrary.
Then, for every fixed realization of $\bH$ and~$\bJ$, %the mutual information $I(\bmy;\bms)$ is lower bounded by 
\begin{align}
    I(\bmy;\bms) \geq \underline{I}(\bH,\bJ)\triangleq \log \big|\inv{\No}\herm{\bU_{\orth}}\bH\herm{\bH}\bU_{\orth} + \bI_{B-I}\big|.
\end{align}
\end{prop}

\begin{prop} \label{prop:unquant_small}
Consider the input-output relation in \eqref{eq:io1}. If $\bmn\sim\setC\setN(\mathbf{0},\No\bI_B)$ is white Gaussian noise and
$\bH\in\opC^{B\times U}$ and $\bJ\in\opC^{B\times I}$ are Rayleigh fading channels, 
\mbox{$\bH\stackrel{\textnormal{i.i.d.}}{\sim}\setC\setN(0,1)$}, $\bJ\stackrel{\textnormal{i.i.d.}}{\sim}\setC\setN(0,1)$, 
% then there exists a linear transform $\mathbb{T}:\opC^{B}\to\opC^{B-I}$ that depends only on $\bJ$ such that 
then the filtered input-output relation
\begin{align}
     \herm{\bU_{\orth}}\bmy = \herm{\bU_{\orth}}(\bH\bms+\bJ\bmw+\bmn)
\end{align}
is equivalent to 
\begin{align}
    \tilde{\bmy}= \tilde{\bH}\bms + \tilde{\bmn},
\end{align}
where $\tilde\bH\in\opC^{(B-I)\times U}$ is Rayleigh fading, 
$\tilde{\bH}\stackrel{\textnormal{i.i.d.}}{\sim}\setC\setN(0,1)$,
and $\tilde\bmn\sim\setC\setN(\mathbf{0}, \No\bI_{B-I})$ is white Gaussian noise.
In other words, any impact of the jammer can be mitigated via spatial filtering by sacrificing 
the equivalent of $I$ antennas at the BS.
\end{prop}

Note that the results in \fref{prop:unquant_mi} and \fref{prop:unquant_small} are independent of the 
distribution of the jammer's transmit signal $\bmw$ (and, in particular, of its power).
In contrast to finite-resolution MIMO communication systems, \emph{infinite-resolution} systems 
are thus resilient to arbitrarily strong single-antenna jammers. 
It is instructive to compare the lower bound $\underline{I}(\bH,\bJ)$ from \fref{prop:unquant_mi}
to the mutual information of a jammer-free $B\times U$ MIMO system when the input is
$\bms\sim\setC\setN(\mathbf{0},\bI_U)$ distributed, which~is
\begin{align}
    I_{\textrm{JF}}(\bH) = \log \big| \inv{\No}\bH\herm{\bH}+\bI_B\big|. \label{eq:jf_mi}
\end{align}

\begin{figure}[tp]
	\centering
    \vspace{-2mm}
    \subfigure[Rayleigh]{
        \hspace{-2mm}
        \includegraphics[height=3.4cm]{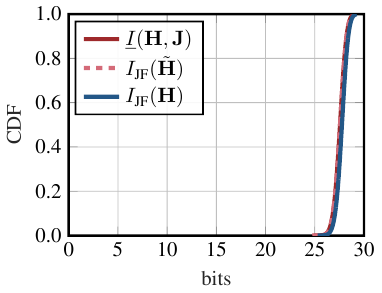}
	    \label{fig:unquantized_rayleigh}
     }
     \hspace{-4mm}
	\subfigure[LoS]{
    	\includegraphics[height=3.4cm]{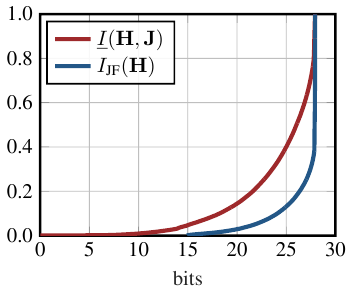}
    	\label{fig:unquantized_los}
     \hspace{-2mm}
    }
    \vspace{-0.1cm}
    \caption{Cumulative distribution functions (CDFs) over channel realizations of the lower bound from \fref{prop:unquant_mi}
    and of the jammer-free mutual information from \eqref{eq:jf_mi} for two different channel models.
    \fref{fig:unquantized_rayleigh} also plots the CDF of a jammer-free $(B\!-\!I)\times U$ MIMO system as described in 
    \fref{prop:unquant_small}.
    }
    \vspace{-2mm}
	\label{fig:unquantized}
\end{figure}

In \fref{fig:unquantized}, we plot the CDFs of $\underline{I}(\bH,\bJ)$ and $I_{\textrm{JF}}(\bH)$
for the same system parameters as in \fref{sec:numeric}. 
For the Rayleigh fading scenario in \fref{fig:unquantized_rayleigh}, we also plot the CDF of $I_{\textrm{JF}}(\tilde{\bH})$,
where $\tilde{\bH}$ is as described in \fref{prop:unquant_small}.
The results show that, especially for Rayleigh fading (\fref{fig:unquantized_rayleigh}), 
$I_{\textrm{JF}}(\bH)$ is only marginally 
lower than $\underline{I}(\bH,\bJ)$. Since the mutual information $I(\bmy;\bms)$ is lower-bounded by $I_{\textrm{JF}}(\bH)$, 
this implies that the achievable communication rates in the jammed system are not significantly lower than those in an equivalent
jammer-free system---irrespective of the strength of the jammer signal.\footnote{
In the case of LoS (\fref{fig:unquantized_los}), the gap between the jammed system and the jammer-free system is larger
than in Rayleigh fading because of the limited spatial diversity. However, also in this case, the lower bound 
on the mutual information is independent of the jammer power.
}
Note also that, as expected from \fref{prop:unquant_small}, the CDF of $\underline{I}(\bH,\bJ)$ in \fref{fig:unquantized_rayleigh} 
overlaps perfectly with the one of $I_{\textrm{JF}}(\tilde{\bH})$.

\section{Proofs} \label{app:proofs}

\subsection{Proof of \fref{thm:mutual_inf}}
We start by defining the sum of the receive interference and the noise, $\bmd\triangleq \bmz+\bmn$, 
so that $\bmy=\bmr+\bmd$ and $\bmq=\mathcal{Q}(\bmy)$.  
Then
\begin{align}
	I(\bmq; \bms)  
	&\leq I(\bmq;\bmr) 
	\leq I(\bmq, \bmd; \bmr) \label{eq:dp_ineq} \\
	&= \underbrace{I(\bmd;\bmr)}_{~=0} + I(\bmq;\bmr|\bmd) \label{eq:mi_dr}
	= H(\bmq|\bmd) - H(\bmq|\bmd,\bmr) \\
    &\leq \sum_{c\in\setC} H(q_c|d_c) - H(\bmq|\bmd,\bmr) \label{eq:cond_red_entr} \\
    &= \sum_{c\in\setC} \big( \underbrace{H(q_c|d_c) - H(q_c|d_c,r_c)}_{\qquad\qquad\!=I(q_c;r_c|d_c)} \big) \label{eq:entropies_are_zero}
\end{align}
where \eqref{eq:dp_ineq} follows from the data processing inequality \cite{cover06a};
where $I(\bmd;\bmr)\!=\!0$ in \eqref{eq:mi_dr} follows from the independence of $\bmd$ and $\bmr$; 
where \eqref{eq:cond_red_entr} follows since conditioning reduces entropy; 
and where \eqref{eq:entropies_are_zero} follows because $H(\bmq|\bmd,\bmr)$ and the $H(q_c|d_c,r_c)$ are all equal to zero.
The second argument in the minimum~of \eqref{eq:thm1} is due to the fact that the mutual information is also 
bounded by the capacity of an unquantized, jammer-free system.
%This concludes the proof.
\hfill$\blacksquare$

\subsection{Proof of \fref{thm:quant}}
We start by defining $d_c \triangleq z_c + n_c$ and the random variable
\begin{align}
	f&\triangleq \mathbbm{1}\big\{ Q_c(r_c+d_c)\neq Q_c(d_c) \big\} \\
	&= \mathbbm{1}\big\{ q_c\neq Q_c(d_c) \big\}. \label{eq:f_deterministic_func}
\intertext{
We then have
}
	I(q_c;r_c|d_c) &=  H(q_c|d_c) - \underbrace{H(q_c|d_c,r_c)}_{~=0} \label{eq:cond_entropy_zero} \\ 
	&= H(q_c|d_c) + \,\overbrace{H(f| q_c,d_c)} \label{eq:f_deterministic} \\
	&= H(q_c,f|d_c)
	= H(f|d_c) + H(q_c|d_c,f) \label{eq:entropy_chain2} \\
	&\leq H(f) + H(q_c|d_c,f) \label{eq:f_cond_red_entr}
\end{align}
where \eqref{eq:cond_entropy_zero} follows from $q_c=Q_c(d_c+r_c)$;
\eqref{eq:f_deterministic} follows because $f$ is a function of $q_c$ and $d_c$ 
(see \eqref{eq:f_deterministic_func}); 
\eqref{eq:entropy_chain2} uses the chain rule of entropy twice;
and \eqref{eq:f_cond_red_entr} follows because conditioning reduces entropy. 
The second term in \eqref{eq:f_cond_red_entr} can now be bounded~as
\begin{align}
	H(q_c|d_c,f)
	&= \Prob(f=0)\cdot \underbrace{H(q_c|d_c,f=0)}_{=0} \nonumber\\
	&\hphantom{=}~+ \Prob(f=1)\cdot H(q_c|d_c,f=1) \label{eq:cond_f_entr} \\
	&\leq \Prob(f=1) \log (M-1) \label{eq:cardinality_bound}
\end{align}
where \eqref{eq:cond_f_entr} follows because, conditional on $f=0$, the quantizer output $q_c$
is a deterministic function of $d_c$; 
and where \eqref{eq:cardinality_bound} follows since the conditional entropy of $q_c$ is bounded
by the logarithm of the alphabet cardinality minus one (conditional on $f=1$, the alphabet value $Q(d_c)$
has probability zero).

Let $\Gamma = \{\gamma_1,\dots,\gamma_{M-1}\}$ denote the set of boundaries~between the quantization
levels of $Q_c$. Then we define the distance between $x\in\mathbb{R}$ and the nearest 
quantization boundary as 
\begin{align}
	\Delta_{\Gamma}(x) &= \min_{\gamma\in\Gamma} |x - \gamma|,
\end{align}
which allows us to bound the probability $\Prob(f=1|d_c=x)$:
\begin{align}
	\!\!\!\Prob(f=1|d_c=x) &\leq \Prob(|r_c|\geq \!\Delta_{\Gamma}(x) )
	\leq \min\!\Big\{1, \frac{\mathsf{R}_c}{\Delta^2_{\Gamma}(x)} \!\Big\},\!\! \label{eq:markov}
\end{align}
where we have used the Markov bound in the second step.
Using the law of total probability, we then bound $\Prob(f=1)$~as
% \\\par
\begin{align}
    \Prob(f=1) &= \int_{\mathbb{R}} \Prob(f=1|d_c=x) ~ p_{d_c}(x) \,dx \\
    &\leq \max_{\substack{\Gamma\subset \mathbb{R},\\ |\Gamma|=M-1}} \int_\mathbb{R} \min\Big\{\frac{\mathsf{R}_c}{\Delta^2_{\Gamma}(x)}, 1 \Big\} ~ p_{d_c}(x) \,dx \label{eq:max_Gamma}\\
    &= \max_{\Gamma} \int_\mathbb{R} \max_{\gamma\in \Gamma}\, g(x-\gamma) ~ p_{d_c}(x) \,dx\text{,} \label{eq:rewrite_min_max}
\end{align}
where $p_{d_c}$ is the density of $d_c\sim\setN(0,(\mathsf{Z}_c+\No)/2)$ and
\begin{align}
    g(x) &\triangleq \min \Big\{1, \frac{\mathsf{R}_c}{|x|^2} \Big\}.
\end{align}
Here, \eqref{eq:max_Gamma} follows from \eqref{eq:markov} and by maximizing over all possible boundary sets $\Gamma$.
Both $g$ and $p_{d_c}$ are positive even Lipschitz continuous functions that are non-increasing on~$[0,\infty).$  

The rest of the proof, which is somewhat technical, serves to show that \eqref{eq:rewrite_min_max}, 
and hence $\Prob(f=1)$, is upper bounded by
\begin{align}
    &\int_\mathbb{R} g\Big(\frac{x}{M-1}\Big)~p_{d_c}(x) \,dx \label{eq:g_spread} \\
    &=\int_{x\in[-(M-1)\sqrt{\mathsf{R}_c},~(M-1)\sqrt{\mathsf{R}_c}]} p_{d_c}(x) \,dx \nonumber \\
	&~~+ \int_{x\notin[-(M-1)\sqrt{\mathsf{R}_c},~(M-1)\sqrt{\mathsf{R}_c}]} \!\!\frac{\mathsf{R}_c(M-1)^2\!}{x^2} \,p_{d_c}(x) \,dx \\
	&= \erf\bigg(\! \frac{(M-1)\sqrt{\mathsf{R}_c}}{\sqrt{2(\mathsf{Z}_c+\No)}} \bigg) \nonumber\\
	&~~+ \sqrt{\frac{2\mathsf{R}_c}{\pi(\mathsf{Z}_c+\No)}}(M-1)\exp\bigg(\!\!-\frac{(M-1)^2\mathsf{SINR}_c}{2}\!\bigg) \nonumber\\
	&~~- (M-1)^2 \frac{\mathsf{R}_c}{\mathsf{Z}_c+\No} \erfc\bigg(\! \frac{(M-1)\sqrt{\mathsf{R}_c}}{\sqrt{2(\mathsf{Z}_c+\No)}} \bigg) \\
	&= \bar{f}(M,\mathsf{SINR}_c), \label{eq:prob_f_bound}
\end{align}
where the last step follows since $\mathsf{SINR}_c=\mathsf{R}_c/(\mathsf{Z}_c+\No)$.
By then combining \eqref{eq:f_cond_red_entr}, \eqref{eq:cardinality_bound}, and \eqref{eq:prob_f_bound}, 
we obtain
\begin{align}
	&I(q_c;r_c|d_c) \nonumber \\
	&\leq H(f) + H(q_c|d_c,f) \\
	&\leq H(f) + \Prob(f=1) \log (M-1) \\
	&\leq H_b(\Prob(f=1)) + \Prob(f=1) \log (M-1) \\
	&\leq \bar{H}_b(\bar{f}(M,\!\mathsf{SINR}_c)) \!+\! \bar{f}(M,\!\mathsf{SINR}_c) \log (M\!-\!1).\hspace{-2mm}
\end{align}
and the theorem follows.

\begin{figure}[tp]
	\centering
	\includegraphics[width=0.96\columnwidth]{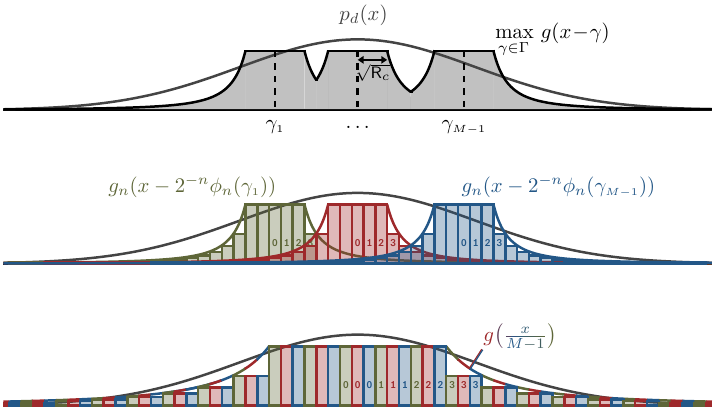}
    % \vspace{-0.2cm}
	\caption{Top: Illustration of the integral in \eqref{eq:rewrite_min_max} for an arbitrary boundary set~$\Gamma$.
    Middle: Illustration of the discretization $g_n\approx g$ (for finite $n$) used in \eqref{eq:insert_limit}. 
    The discretization $p_n\approx p_{d_c}$ is not shown to preserve figure legibility.
    Bottom: For any $n$ and $\Gamma$, the inner product of $p_{d_c}$ (or its discretized version $p_n$) with 
    $\max_{\gamma\in\Gamma} g_n(x-2^{-n}\phi_n(\gamma))$ is upper bounded by rearranging the piece-wise constant segments
    of the $M\!-\!1$ translations of $g_n$ as shown in the bottom figure.
    The rearranged segments converge to $g(x/(M\!-\!1))$ as $n\to\infty$.
    }
    \vspace{-0.1cm}
	\label{fig:proof_sketch}
\end{figure}

To fill the gap and show that \eqref{eq:g_spread} bounds \eqref{eq:rewrite_min_max} (see also \fref{fig:proof_sketch}),
we write \eqref{eq:rewrite_min_max} as the limit of a sequence of 
integrals~over the product of piecewise constant functions $g_n\!\approx\! g$ and $p_n\!\approx\! p_{d_c}$. 
Specifically, for $n=1,2,\dots$ and $\ell\in\mathbb{Z}$ we define the intervals
\begin{align}
    \!\!R_n(\ell) &\triangleq [2^{-n}\ell,2^{-n}\ell+2^{-n})
\end{align}
and the functions 
\begin{align}
    % \!\!\phi_n(x) &\triangleq \min\big\{m\in2^{-n}\mathbb{Z}\big|x\leq m\big\},\\
    \!\!\phi_n(x) &\triangleq \lceil2^n x\rceil,\\
    g_n(x) &\triangleq \sum_{\ell\in\mathbb{Z}}
    \Big(\inf_{R_n(\ell)}g\Big) \; \mathbbm{1}\{x\in R_n(\ell)\}, \label{eq:define_gn} \\ 
    p_n(x) &\triangleq\sum_{\ell\in\mathbb{Z}}
    \Big(\inf_{R_n(\ell)}p_{d_c}\Big) \; \mathbbm{1}\{x\in R_n(\ell)\}. 
\end{align}
For all $x,a\in\mathbb{R}$, it holds that
\begin{align}
    g\big(x-a\big)&=\lim_{n\rightarrow\infty}g_n\big(x-2^{-n}\phi_n(a)\big), \label{eq:lim_g}
\end{align}
since, for $n=1,2,\dots$, we have
\begin{align}
    &\big|g_n\big(x-2^{-n}\phi_n(a)\big)-g\big(x-a\big)\big| \nonumber \\
    % =&\Big| g_n\big(x-\phi_n(a)\big)-g\big(x-\phi_n(a)\big)\nonumber\\+&\;g\big(x-\phi_n(a)\big)-g\big(x-a\big) \Big| \label{eq:add_with_zero}\\
    &\leq \big|g_n\left(x-2^{-n}\phi_n(a)\right)\!-\!g\left(x-2^{-n}\phi_n(a)\right)\big| \nonumber \\
    &~~+ \big|g\left(x-2^{-n}\phi_n(a)\right)\!-\!g\left(x\!-\!a\right)\big| \label{eq:triangle inequality} \\
    &< L 2^{-n} + \big|g\left(x-2^{-n}\phi_n(a)\right)\!-\!g\left(x\!-\!a\right)\big| \label{eq:lipschitz_and_def} \\
    &< 2L \,2^{-n},\label{eq:lipschitz_and_def2}
\end{align}
where $L$ is the Lipschitz constant of $g$.
Here, \eqref{eq:triangle inequality} follows from the triangle inequality;
\eqref{eq:lipschitz_and_def} follows from the $L$-Lipschitz continuity of $g$
and the fact that the length of the interval $R_n(\ell)$ in \eqref{eq:define_gn} is bounded by $2^{-n}$; 
and \eqref{eq:lipschitz_and_def2} follows from the $L$-Lipschitz continuity of $g$ and the fact that \mbox{$|2^{-n}\phi_n(a)-a|<2^{-n}$.}

An analogous line of reasoning shows that, for all $x\in\mathbb{R}$,
\begin{align}
    p_{d_c}\big(x)&=\lim_{n\rightarrow\infty}p_n\big(x\big).\label{eq:lim_p}
\end{align}

\begin{figure*}[tp]
  \normalsize
\begin{align}
    % &\!\!\Prob(f=1)
    % \leq
    \max_{\Gamma} 
    \int_\mathbb{R} 
    \max_{\gamma\in \Gamma}\,
    &g(x-\gamma) ~ p_{d_c}(x) 
    \,dx 
    % \nonumber
    % \\
    % &=
    =
    \max_{\Gamma} 
    \int_\mathbb{R} 
    \lim_{n\rightarrow\infty}
    \max_{\gamma\in \Gamma} \,
    g_n\big(x-2^{-n}\phi_n(\gamma)\big) ~ p_n\big(x\big) 
    \,dx
    \label{eq:insert_limit}
    \\
    &=
    \lim_{n\rightarrow\infty}
    \max_{\Gamma} 
    \int_\mathbb{R} 
    \max_{\gamma\in \Gamma} 
    g_n\big(x-2^{-n}\phi_n(\gamma)\big) ~ p_n\big(x\big) 
    \,dx
    \label{eq:dominated_conv}
    \\
    &=
    \lim_{n\rightarrow\infty}
    \max_{\Gamma} 
    \int_\mathbb{R} 
    \max_{\gamma\in \Gamma}\,
    ~\sum_{\ell\in\mathbb{Z}} 
    ~
    \sum_{\ell'\in\mathbb{Z}} 
    \Big(\inf_{R_n(\ell)}g\Big)
    \Big(\inf_{R_n(\ell')}p_{d_c}\Big)
    \mathbbm{1}\{x-2^{-n}\phi_n(\gamma)\in R_n(\ell)\}
    ~\mathbbm{1}\{x\in R_n(\ell')\}
    \,dx \label{eq:ultra_long_equation}
    \\
    &=
    \lim_{n\rightarrow\infty}
    \max_{\Gamma} 
    \int_\mathbb{R} 
    \max_{\gamma\in \Gamma}\, 
    \sum_{\ell\in\mathbb{Z}} 
    \Big(\inf_{R_n(\ell)}g\Big)
    \Big(\inf_{R_n(\ell+\phi_n(\gamma))}p_{d_c}\Big)
    \mathbbm{1}\{x\in R_n(\ell+\phi_n(\gamma))\}
    \,dx, \label{eq:simplified_ultra_long_equation}
    \\
    &\leq 
    \lim_{n\rightarrow\infty}
    \max_{\pi_1,\dots,\pi_{M-1}} 
    \int_\mathbb{R} 
    \max_{k\in[M-1]}\, 
    \sum_{\ell\in\mathbb{Z}} 
    \Big(\inf_{R_n(\ell)}g\Big)
    \Big(\inf_{R_n(\pi_k(\ell))}p_{d_c}\Big) 
    ~ \mathbbm{1}\{x\in R_n(\pi_k(\ell))\}
    \,dx \label{eq:new_relaxation} \\
    &= 
    \lim_{n\rightarrow\infty}
    \max_{\boldsymbol{\pi}\in\Pi} 
    \int_\mathbb{R} 
    \max_{k\in[M-1]}\, 
    \sum_{\ell\in\mathbb{Z}} 
    \Big(\inf_{R_n(\ell)}g\Big)
    \Big(\inf_{R_n(\pi_k(\ell))}p_{d_c}\Big) 
    ~ \mathbbm{1}\{x\in R_n(\pi_k(\ell))\}
    \,dx \label{eq:upper_bound_can_be_obtained} \\
    &= 
    \lim_{n\rightarrow\infty}
    \max_{\boldsymbol{\pi}\in\Pi}
    \int_\mathbb{R} 
    \sum_{k\in[M-1]}\, 
    \sum_{\ell\in\mathbb{Z} } 
    \Big(\inf_{R_n(\ell)}g\Big)
    \Big(\inf_{R_n(\pi_k(\ell))}p_{d_c}\Big) 
    ~ \mathbbm{1}\{x\in R_n(\pi_k(\ell))\}
    \,dx \label{eq:new_max_to_sum} \\
    &= 
    \lim_{n\rightarrow\infty}
    \max_{\boldsymbol{\pi}\in\Pi} 
    \sum_{k\in[M-1]}\, 
    \sum_{\ell\in\mathbb{Z} } 
    \Big(\inf_{R_n(\ell)}g\Big)
    \Big(\inf_{R_n(\pi_k(\ell))}p_{d_c}\Big) 
    \underbrace{
    \int_\mathbb{R} 
    ~ \mathbbm{1}\{x\in R_n(\pi_k(\ell))\}
    \,dx
    }_{\quad~\,=2^{-n}} 
    \label{eq:take_sum_out_of_int}\\
    &= 
    \lim_{n\rightarrow\infty}
    \int_\mathbb{R} 
    \sum_{(k,\ell)\in[M-1]\times\mathbb{Z}}
    \Big(\inf_{R_n(\ell)}g\Big)
    \Big(\inf_{R_n\big(k-1+\ell(M-1)\big)}p_{d_c}\Big) 
    ~ \mathbbm{1}\Big\{x\in R_n\big(k-1+\ell(M-1)\big)\Big\}
    \,dx \label{eq:insert_pi_hat} \\
    &= 
    \lim_{n\rightarrow\infty}
    \int_\mathbb{R} 
    \sum_{\ell'\in\mathbb{Z}} 
    \Big(\inf_{R_n(\lfloor\ell'/(M-1)\rfloor)}g\Big)
    \Big(\inf_{R_n(\ell')}p_{d_c}\Big) 
    ~ \mathbbm{1}\{x\in R_n(\ell')\}
    \,dx \label{eq:invert_pi} 
    \\
    &=
    \lim_{n\rightarrow\infty}
    \int_\mathbb{R} 
    \sum_{\ell\in\mathbb{Z}} 
    \Big(\inf_{R_n(\lfloor\ell/(M-1)\rfloor)}g\Big)
    \mathbbm{1}\{x\in R_n(\ell)\}
    \sum_{\ell'\in\mathbb{Z}}  
    \Big(\inf_{R_n(\ell')}p_{d_c}\Big) 
    \mathbbm{1}\{x\in R_n(\ell')\}
    \,dx 
    \label{eq:make_two_sums_again}
    \\
    &=
    \lim_{n\rightarrow\infty}
    \int_\mathbb{R} 
    \sum_{\ell\in\mathbb{Z}} 
    \Big(\inf_{R_n(\ell)}g\Big)
    \mathbbm{1}\{x/(M-1)\in R_n(\ell)\}
    \sum_{\ell'\in\mathbb{Z}}  
    \Big(\inf_{R_n(\ell')}p_{d_c}\Big) 
    \mathbbm{1}\{x\in R_n(\ell')\}
    \,dx
    \label{eq:from_cond_to_argument}
    \\
    &=
    \int_\mathbb{R} 
    g\Big(\frac{x}{M-1}\Big)~p_{d_c}(x)
    \,dx, 
    \label{eq:take_lim_inside_again}
\end{align}
  \hrulefill
\end{figure*}

\noindent
Plugging these limits into \eqref{eq:rewrite_min_max}, and using the product law for limits and
the fact that the maximum is a continuous function, gives \eqref{eq:insert_limit};
\eqref{eq:dominated_conv} can be shown by applying Lebesgue's dominated convergence theorem;
and \eqref{eq:simplified_ultra_long_equation} follows (i) from the fact that $R_n(\ell)\cap R_n(\ell')=\emptyset$ 
for all $n$ and all $\ell\neq\ell'$, and (ii) from the fact that $\text{image}\left(\phi_n\right)\subset \mathbb{Z}$. 
The step to \eqref{eq:new_relaxation} follows by relaxing the maximization problem over boundary sets $\Gamma$ 
to an optimization over surjections $\pi_k:\mathbb{Z}\twoheadrightarrow \mathbb{Z}, ~k\in[M-1]$. 
We then define $\Pi$ as the set of $(M\!-\!1)$-tuples $\boldsymbol{\pi}=(\pi_1,\dots,\pi_{M-1})$ 
of surjections $\pi_k$ for which, for every two surjections $\pi_k, \pi_{k'}$ out of $\boldsymbol{\pi}$, 
it holds that $\textit{image}(\pi_{k'})\cap\textit{image}(\pi_k)=\emptyset$.
Since both $g$ and $p_d$ are positive, the maximum in \eqref{eq:new_relaxation} can be attained by an $(M\!-\!1)$-tuple
of surjections that is contained in $\Pi$ (overlapping sections do not help, see \fref{fig:proof_sketch}), which gives \eqref{eq:upper_bound_can_be_obtained}.
For such non-overlapping surjections, we have equality between \eqref{eq:upper_bound_can_be_obtained} and 
\eqref{eq:new_max_to_sum}; the step to \eqref{eq:take_sum_out_of_int} follows from the monotone convergence theorem.
Since $g$ and $p_{d_c}$ are non-negative even functions that are non-increasing on $[0,\infty)$, 
we can deduce that optimal surjections $\boldsymbol{\hat\pi}=(\hat\pi_1, \dots, \hat\pi_{M-1})$ are given by
\begin{align}
    \hat{\pi}_k(\ell) = k - 1 + \ell(M - 1), \qquad k=1,\dots,M-1. \nonumber
\end{align}
Inserting these into \eqref{eq:take_sum_out_of_int} gives \eqref{eq:insert_pi_hat}. 
The step to \eqref{eq:invert_pi} is obtained by interpreting $\boldsymbol{\hat\pi}$ as 
a bijection from $[M-1]\times\mathbb{Z}$ to $\mathbb{Z}$, where \eqref{eq:insert_pi_hat} sums over the domain 
of $\boldsymbol{\hat\pi}$ and \eqref{eq:invert_pi} sums over the image of $\boldsymbol{\hat\pi}$.
\eqref{eq:make_two_sums_again} follows from the fact that
$\sum_{\ell}\mathbbm{1}\{x\in R_n(\ell)\} = \sum_{\ell}\sum_{\ell'}\mathbbm{1}\{x\in R_n(\ell)\} \mathbbm{1}\{x\in R_n(\ell')\}$.
In \eqref{eq:from_cond_to_argument}, we use that $\lceil\ell/(M\!-\!1)\rceil$ stays constant for $M\!-\!1$ consecutive values of~$\ell$.
Finally, \eqref{eq:take_lim_inside_again} follows by taking the limit inside the integral 
(which is warranted by Lebesgue's dominated convergence theorem) and using \eqref{eq:lim_g} and \eqref{eq:lim_p}.
This concludes the proof. 
\hfill$\blacksquare$

\subsection{Proof of \fref{thm:main}}
We first show that the function $\bar{I}(M,\rho,\bH,\bJ)$ in \eqref{eq:sinr_bound} is an increasing function of $\mathsf{SINR}_c$.
This follows since the function $x\mapsto \min\{\log M, \bar{H}_b(x) + x \log(M\!-\!1)\}$ is increasing, 
and since the function $\bar{f}(M,\mathsf{SINR}_c)$ in \eqref{eq:f_bound} is increasing in $\mathsf{SINR}_c$
(this can be seen by taking the derivative with respect to $\mathsf{SINR}_c$ and observing that it is non-negative).
Thus, the result follows by proving 
that $\overline{\mathsf{SINR}}_c$ from \eqref{eq:sinr_bound} is 
an upper bound to $\mathsf{SINR}_c$ from \fref{thm:quant}
for suitable marginal distributions of the jammer interference $z_c$, the noise $n_c$, and the legitimate signal~$r_c$:

Consider an index $c\in\{(b,\mathfrak{r}),\,(b,\mathfrak{i})\}$. 
Since the entries $s_u$ of $\bms$ are uncorrelated and satisfy $\Ex{}{|s_u|^2}\leq 1$, it follows\footnote{If
$\bms\sim\setC\setN(\mathbf{0},\bI_U)$, then 
$r_{(b,\mathfrak{r})} + i r_{(b,\mathfrak{i})} = \bmh_{(b)}\bms \sim \setC\setN(0,\|\bmh_{(b)}\|_2^2)$, 
so that $r_{(b,\mathfrak{r})}$ and $r_{(b,\mathfrak{i})}$ are $\setN(\mathbf{0},\|\bmh_{(b)}\|_2^2/2)$ distributed. 
This implies \fref{rem:sinr}.
}
\begin{align}
    \Ex{}{|r_c|^2} &\leq \Ex{}{|r_{(b,\mathfrak{r})} + i r_{(b,\mathfrak{i})}|^2} \\
    &= \Ex{}{|\tp{\bmh_{(b)}}\bms|^2} 
    = \tp{\bmh_{(b)}} \Ex{}{\bms\herm{\bms}} \bmh_{(b)}^\ast \\
    &= \tp{\bmh_{(b)}} \text{diag}(\Ex{}{|s_1|^2},\dots,\Ex{}{|s_U|^2}) \bmh_{(b)}^\ast \\
    &\leq \|\bmh_{(b)}\|_2^2.
\end{align}
Since $\bmw\sim\setC\setN(\mathbf{0},\rho\bI_I)$, the distribution of $z_{(b,\mathfrak{r})} + i z_{(b,\mathfrak{i})}$ is 
\begin{align}
    z_{(b,\mathfrak{r})} + i z_{(b,\mathfrak{i})} = \bmj_{(b)}\bmw \sim \setC\setN(\mathbf{0},\rho\|\bmj\|_2^2), 
\end{align}
meaning that both $z_{(b,\mathfrak{r})}$ and $z_{(b,\mathfrak{i})}$ are $\setN(\mathbf{0},\rho\|\bmj\|_2^2/2)$ distributed.
And since $\bmn\sim\setC\setN(\mathbf{0},\No\bI_I)$, the distribution of both $n_{(b,\mathfrak{r})}$ and $n_{(b,\mathfrak{i})}$
is $\setN(0,\No/2)$. Hence, the conditions of \fref{thm:quant} are fulfilled with a SINR that is bounded by
\begin{align}
    \mathsf{SINR}_c \leq \frac{\|\bmh_{(b)}\|_2^2}{\rho\|\bmj_{(b)}\|_2^2/2 + \No/2} = \overline{\mathsf{SINR}}_c.
\end{align}
From this, the result follows. 
\hfill$\blacksquare$

\subsection{Proof of \fref{prop:vanish}}
Since all rows $\bmj_{(b)}$ of $\bJ$ are nonzero, we have $\overline{\mathsf{SINR}}_c\xrightarrow{\rho\to\infty} 0$ 
for all $c\in\setC$, see \eqref{eq:sinr_bound}.
The result now follows directly from \fref{thm:main}, considering the fact that
$\bar{f}(M,\mathsf{SINR}_c)\xrightarrow{\mathsf{SINR}_c\to0}0$ and that $\bar{H}_b(x)\xrightarrow{x\to0}0$.
\hfill$\blacksquare$

\subsection{Proof of \fref{prop:6dB}} \label{app:proof_6db}
Since all terms of $\bar{I}(M,\rho,\bH,\bJ)$ in \eqref{eq:mutual_inf_mainbound} depend on $\rho$ and $M$ and 
have the same structure, it suffices to consider the conditions under which one of the terms tends to 
zero as $\rho\to\infty$. Let $M=M(\rho)$ be a function of the jammer power~$\rho$, and recall that 
$\overline{\mathsf{SINR}}_c=\overline{\mathsf{SINR}}_c(\rho)$ is a function of $\rho$, too (see~\eqref{eq:sinr_bound}).
By dropping the minimum from \eqref{eq:cond_entropy_bound}, the $c$-th term of the bound in \eqref{eq:mutual_inf_mainbound} 
can then be upper-bounded with
\begin{align}
	T_c(\rho)\triangleq \bar{H}_b\big(\bar{f}(\rho)\big) + \bar{f}(\rho)\log\big(M(\rho)-1\big), \label{eq:termbound}
\end{align}
where $\bar{f}(\rho)\triangleq \bar{f}\big(M(\rho),\overline{\mathsf{SINR}}_c(\rho)\big)$ is defined on $\rho\in[0,\infty)$.
Both terms in \eqref{eq:termbound} are nonnegative, 
so that $T_c(\rho)$ stays bounded away from zero if and only if either of its terms stays bounded away from zero. 
The first term stays bounded away from zero if and only if $\bar{f}(\rho)$ does;
the second term stays bounded away from zero if and only if $\bar{f}(\rho)\log\big(M(\rho)\!-\!1\big)$ does. 
Since $M(\rho)\geq2$ holds, $\bar{f}(\rho)\log\big(M(\rho)\!-\!1\big)\to0$ implies \mbox{$\bar{f}(\rho)\to0$},
so we only need to consider the conditions under which $\bar{f}(\rho)\log\big(M(\rho)\!-\!1\big)$ stays bounded 
away from zero as~$\rho\to\infty$.

We start by providing the following bound on $\bar{f}(\rho)$, see~\eqref{eq:f_bound}:
\begin{align}
    0 \leq \bar{f}(\rho) 
    &\leq \erf\!\bigg(\!\frac{(M(\rho)\!-\!1)\sqrt{\overline{\mathsf{SINR}}_c(\rho)}}{\sqrt{2}}\bigg) \nonumber \\
    &~~+ \sqrt{\frac{2}{\pi}}(M(\rho)\!-\!1)\sqrt{\overline{\mathsf{SINR}}_c(\rho)}    \label{eq:fbarbar_1} \\
    &\leq 2\sqrt{\frac{2}{\pi}}(M(\rho)\!-\!1)\sqrt{\overline{\mathsf{SINR}}_c(\rho)} \label{eq:fbarbar_2} \\
    &= \bar{\bar{f}}\big(M(\rho),\overline{\mathsf{SINR}}_c(\rho)\big). \label{eq:derive_fbarbar}
\end{align}
In \eqref{eq:fbarbar_1} we have used that $\exp(-x)\leq1$ and $\erfc(\sqrt{x})\leq0$ for $x\geq0$; 
in \eqref{eq:fbarbar_2}, we have used that\footnote{This follows by
$\erf(x)=\!\frac{2}{\sqrt{\pi}}\!\int_0^x e^{-t^2} dt \leq\! \frac{2}{\sqrt{\pi}}\! \int_0^x \!1\, dt =\! \frac{2x}{\sqrt{\pi}}$
for $x\!\geq\!0$.}
$\erf(x)\leq 2x/\sqrt{\pi}$ for $x\geq0$.
Equations \eqref{eq:fbarbar_1}\,--\,\eqref{eq:derive_fbarbar} also imply \fref{rem:simplified}.
Since $\sqrt{8/\pi}\leq2$, $M(\rho)-1\leq M(\rho)$, and 
$\overline{\mathsf{SINR}}_c(\rho)\leq \frac{2\|\bmh_{(b)}\|_2^2}{\rho\|\bmj_{(b)}\|_2^2}$,
it follows that 
\begin{align}
    \!\!\!0 \leq& \bar{f}(\rho)\log\!\big(M(\rho)\!-\!1\big)\! 
     \leq\!  \frac{\sqrt{32}\|\bmh_{(b)}\|_2}{\|\bmj_{(b)}\|_2\!} \frac{M(\rho)\!}{\sqrt{\rho}} \log\!\big(M(\rho)\big).\!\!\!
     \label{eq:sandwich}
\end{align}
For any $\epsilon>0$, if $M(\rho)=\rho^{\frac{1}{2}-\epsilon}$, then the right-hand-side of \eqref{eq:sandwich} tends to zero as 
$\rho\to\infty$, which implies that $\bar{f}(\rho)\log\!\big(M(\rho)\!-\!1\big)$ tends to zero as well. 

It follows that $M\prec \sqrt{\rho}$ implies \mbox{$\lim_{\rho\to\infty}\bar{I}(M,\rho,\bH,\bJ)=0$}
and, logically equivalently, that $\lim_{\rho\to\infty}\bar{I}(M,\rho,\bH,\bJ)>0$ implies  $M\succeq \sqrt{\rho}$.
\hfill$\blacksquare$

\subsection{Proof of \fref{prop:unquant_mi}}
We have
\begin{align}
    I(\bmy;\bms) &\geq I(\herm{\bU_{\orth}}\bmy;\bms)
    = h(\herm{\bU_{\orth}}\bmy) - h(\herm{\bU_{\orth}}\bmy|\bms) \label{eq:ys_dp} \\
    &= \log\big| \pi e \herm{\bU_{\orth}}(\bH\herm{\bH} + \No\bI_B)\bU_{\orth} \big| - \log\big| \pi e \No \bI_{B-I} \big| 
    \label{eq:bigstep}\\
    &= \log\big| \inv{\No}\herm{\bU_{\orth}}\bH\herm{\bH}\bU_{\orth} + \bI_{B-I}\big|,
\end{align}    
where \eqref{eq:ys_dp} follows from the data-processing inequality;
and \eqref{eq:bigstep} follows since
$\bU_{\orth}\bmy\sim\setC\setN(\mathbf{0},\herm{\bU_{\orth}}(\bH\herm{\bH} + \No\bI_B)\bU_{\orth})$ 
(because $\herm{\bU_{\orth}}\bJ=\mathbf{0}$), since, conditional on $\bms$, the vector $\bU_{\orth}\bmy$
is circularly-symmetric complex Gaussian with covariance matrix $\No \bU_{\orth}\herm{\bU_{\orth}}=\No \bI_{B-I}$, 
and since the differential entropy of a complex Gaussian with covariance matrix $\mathbf{\Sigma}$ is
$\log|\pi e \mathbf{\Sigma}|$.
\hfill$\blacksquare$

\subsection{Proof of \fref{prop:unquant_small}}
We have
\begin{align}
    \herm{\bU_{\orth}}\bmy 
    &= \underbrace{\herm{\bU_{\orth}}\bH}_{\triangleq\tilde{\bH}}\bms 
    + \underbrace{\herm{\bU_{\orth}}\bJ}_{=0} \bmw + \underbrace{\herm{\bU_{\orth}}\bmn}_{\triangleq\tilde{\bmn}}
    = \tilde{\bH}\bms + \tilde{\bmn}.
\end{align}
Since $\bH\!\stackrel{\textnormal{i.i.d.}}{\sim}\!\setC\setN(0,1)$ and the rows of $\herm{\bU_{\orth}}$ are \mbox{orthonormal,}
it follows that $\tilde{\bH}\!\stackrel{\textnormal{i.i.d.}}{\sim}\!\setC\setN(0,1)$. 
The orthonormality of the rows of $\herm{\bU_{\orth}}$ as well as the fact that 
$\bmn\sim\setC\setN(\mathbf{0},\No\bI_B)$ imply $\tilde{\bmn}\sim\setC\setN(\mathbf{0},\No\bI_{B-I})$.
This concludes the proof.
\hfill$\blacksquare$

\vspace{5mm}
\linespread{1}


\begin{thebibliography}{10}
\providecommand{\url}[1]{#1}
\csname url@samestyle\endcsname
\providecommand{\newblock}{\relax}
\providecommand{\bibinfo}[2]{#2}
\providecommand{\BIBentrySTDinterwordspacing}{\spaceskip=0pt\relax}
\providecommand{\BIBentryALTinterwordstretchfactor}{4}
\providecommand{\BIBentryALTinterwordspacing}{\spaceskip=\fontdimen2\font plus
\BIBentryALTinterwordstretchfactor\fontdimen3\font minus
  \fontdimen4\font\relax}
\providecommand{\BIBforeignlanguage}[2]{{%
\expandafter\ifx\csname l@#1\endcsname\relax
\typeout{** WARNING: IEEEtran.bst: No hyphenation pattern has been}%
\typeout{** loaded for the language `#1'. Using the pattern for}%
\typeout{** the default language instead.}%
\else
\language=\csname l@#1\endcsname
\fi
#2}}
\providecommand{\BIBdecl}{\relax}
\BIBdecl

\bibitem{pirayesh2022jamming}
H.~Pirayesh and H.~Zeng, ``Jamming attacks and anti-jamming strategies in
  wireless networks: A comprehensive survey,'' \emph{{IEEE} Commun. Surveys
  Tuts.}, vol.~9, no.~2, pp. 767--809, 2022.

\bibitem{leost2012interference}
Y.~L{\'e}ost, M.~Abdi, R.~Richter, and M.~Jeschke, ``Interference rejection
  combining in {LTE} networks,'' \emph{Bell Labs Tech.~J.}, vol.~17, no.~1, pp.
  25--50, Jun. 2012.

\bibitem{marti2023universal}
G.~Marti and C.~Studer, ``Universal {MIMO} jammer mitigation via secret
  temporal subspace embeddings,'' in \emph{Proc. Asilomar Conf. Signals, Syst.,
  Comput.}, Oct. 2023, pp. 1--8.

\bibitem{mollen16c}
C.~Moll{\'e}n, J.~Choi, E.~G. Larsson, and R.~W. {Heath Jr.}, ``Uplink
  performance of wideband massive {MIMO} with one-bit {ADCs},'' \emph{{IEEE}
  Trans. Wireless Commun.}, vol.~16, no.~1, pp. 87--100, Jan. 2017.

\bibitem{jacobsson17b}
S.~Jacobsson, G.~Durisi, M.~Coldrey, U.~Gustavsson, and C.~Studer, ``Throughput
  analysis of massive {MIMO} uplink with low-resolution {ADCs},'' \emph{{IEEE}
  Trans. Wireless Commun.}, vol.~16, no.~6, pp. 4038--4051, Jun. 2017.

\bibitem{demir2020bussgang}
O.~T. Demir and E.~Bjornson, ``The {Bussgang} decomposition of nonlinear
  systems: Basic theory and {MIMO} extensions [lecture notes],'' \emph{{IEEE}
  Signal Process. Mag.}, vol.~38, no.~1, pp. 131--136, 2020.

\bibitem{block2006performance}
F.~J. Block, ``Performance of wideband digital receivers in jamming,'' in
  \emph{Proc. IEEE Mil. Commun. Conf. (MILCOM)}, 2006, pp. 1--7.

\bibitem{do18a}
T.~T. {Do}, E.~{Bj\"ornsson}, E.~G. {Larsson}, and S.~M. {Razavizadeh},
  ``Jamming-resistant receivers for the massive {MIMO} uplink,'' \emph{{IEEE}
  Trans. Inf. Forensics Security}, vol.~13, no.~1, pp. 210--223, Jan. 2018.

\bibitem{li17b}
Y.~Li, C.~Tao, G.~Seco-Granados, A.~Mezghani, A.~L. Swindlehurst, and L.~Liu,
  ``Channel estimation and performance analysis of one-bit massive {MIMO}
  systems,'' \emph{{IEEE} Trans. Signal Process.}, vol.~65, no.~15, pp.
  4075--4089, Aug. 2017.

\bibitem{bucheli2024vlsi}
F.~Bucheli, O.~Casta{\~n}eda, G.~Marti, and C.~Studer, ``A jammer-mitigating
  267 {Mb/s} 3.78 mm$^2$ 583 m{W} 32$\times$8 multi-user {MIMO} receiver in
  {22FDX},'' in \emph{Proc. IEEE Int. Symp. VLSI Technol. Circuits}.

\bibitem{lu2016effect}
Z.~Lu, J.~Nie, J.~Li, and G.~Ou, ``Effect of {ADC} non-ideal characteristics on
  {GNSS} antenna array anti-jamming,'' in \emph{Int. Conf. Comput. Sci. Netw.
  Technol. (ICCSNT)}, 2016, pp. 512--516.

\bibitem{pirzadeh2019mitigation}
H.~Pirzadeh, G.~Seco-Granados, and A.~L. Swindlehurst, ``Mitigation of jamming
  attack in massive {MIMO} with one-bit {FBB} sigma-delta {ADCs},'' in
  \emph{Asilomar Conf. Signals, Syst., Comput.}, 2019, pp. 1700--1704.

\bibitem{marti2021hybrid}
G.~Marti, O.~Casta\~neda, S.~Jacobsson, G.~Durisi, T.~Goldstein, and C.~Studer,
  ``Hybrid jammer mitigation for all-digital {mmWave} massive {MU-MIMO},'' in
  \emph{Proc. Asilomar Conf. Signals, Syst., Comput.}, Nov. 2021, pp. 93--99.

\bibitem{marti2021snips}
G.~Marti, O.~Casta\~neda, and C.~Studer, ``Jammer mitigation via beam-slicing
  for low-resolution {mmWave} massive {MU-MIMO},'' \emph{{IEEE} Open J.
  Circuits Syst.}, vol.~2, pp. 820--832, Dec. 2021.

\bibitem{jiang2023ris}
W.~Jiang, K.~Huang, M.~Yi, Y.~Chen, and L.~Jin, ``{RIS}-based reconfigurable
  antenna for anti-jamming communications with bit-limited {ADCs},'' in
  \emph{Proc. IEEE Global Commun. Conf. (GLOBECOM)}, 2023, pp. 1433--1438.

\bibitem{teeti2021one}
M.~A. Teeti, ``One-bit window comparator based jamming detection in massive
  {MIMO} system,'' in \emph{IEEE Veh. Technol. Conf. Spring (VTC-Spring)}, Apr.
  2021, pp. 1--5.

\bibitem{cover06a}
T.~M. Cover and J.~A. Thomas, \emph{Elements of Information Theory}.\hskip 1em
  plus 0.5em minus 0.4em\relax Wiley, 2006.

\end{thebibliography}
\end{document}